\newif\ifpdf
\ifx\pdfoutput\undefined
\pdffalse 
\else
\pdfoutput=1 
\pdftrue \fi

\newif\iffinal
\finalfalse	
\finaltrue 

\documentclass[reqno,twoside,11pt]{amsart}
\iffinal\else\usepackage[notref,notcite]{showkeys}\fi\usepackage{cite}
\usepackage{amsmath}
\usepackage{amsfonts}
\usepackage{amssymb}
\usepackage{wrapfig}
\usepackage{graphicx}
\usepackage{graphics}
\usepackage{verbatim}
\usepackage{xr}
\usepackage{exscale}

\def\version{June 21, 2004}

\IfFileExists{epsf.def}{\input epsf.def}{\usepackage{epsf}}

\IfFileExists{myowntimes.sty}{\usepackage{myowntimes}}
	{\usepackage{times}\usepackage{mathrsfs}}
\DeclareFontFamily{OT1}{eusb}{} \DeclareFontShape{OT1}{eusb}{m}{n}
{<5> <6> <7> <8> <9> <10> <11> <12> <14.4> eusb10}{}
\DeclareMathAlphabet{\eusb}{OT1}{eusb}{m}{n}

\DeclareFontFamily{OT1}{eusm}{} \DeclareFontShape{OT1}{eusm}{m}{n}
{<5> <6> <7> <8> <9> <10> <11> <12> <14.4> eusm10}{}
\DeclareMathAlphabet{\eusm}{OT1}{eusm}{m}{n}

\DeclareFontFamily{OT1}{eufm}{} \DeclareFontShape{OT1}{eufm}{m}{n}
{<5> <6> <7> <8> <9> <10> <11> <12> <14.4> eufm10}{}
\DeclareMathAlphabet{\mathfrak}{OT1}{eufm}{m}{n}

\DeclareFontFamily{OT1}{fraktura}{}
\DeclareFontShape{OT1}{fraktura}{m}{n} {<5> <6> <7> <8> <9> <10>
<11> <12> <13> <14.4> [1.1] eufm10}{}
\DeclareMathAlphabet{\fraktura}{OT1}{fraktura}{m}{n}

\DeclareFontFamily{OT1}{cmfi}{} \DeclareFontShape{OT1}{cmfi}{m}{n}
{<5> <6> <7> <8> <9> <10> <11> <12> <13> <14.4> [0.9] cmfi10}{}
\DeclareMathAlphabet{\cmfi}{OT1}{cmfi}{b}{n}

\DeclareFontFamily{OT1}{cmss}{} \DeclareFontShape{OT1}{cmss}{m}{n}
{<5> <6> <7> <8> <9> <10> <11> <12> <13> <14.4> cmss10}{}
\DeclareMathAlphabet{\cmss}{OT1}{cmss}{m}{n}

\newtheoremstyle{thm}{1.5ex}{1.5ex}{\itshape\rmfamily}{}
{\bfseries\rmfamily}{}{2ex}{}

\newtheoremstyle{def}{1.5ex}{1.5ex}{\rmfamily}{}
{\bfseries\rmfamily}{}{2ex}{}

\newtheoremstyle{rem}{1.3ex}{1.3ex}{\rmfamily}{}
{\itshape}
{} {1.5ex}{}

\newenvironment{proofsect}[1]
{\vskip0.1cm\noindent{\rmfamily\itshape#1.}}{\qed\vspace{0.15cm}}

\theoremstyle{thm}
\newtheorem{theorem}{Theorem}[section]
\newtheorem{lemma}[theorem]{Lemma}
\newtheorem{proposition}[theorem]{Proposition}

\newtheorem*{Main Theorem}{Main Theorem.}
\newtheorem{corollary}[theorem]{Corollary}

\theoremstyle{def}
\newtheorem{definition}{Definition}

\theoremstyle{rem}
\newtheorem{remark}{{\itshape Remark}}[]

\numberwithin{equation}{section}

\renewcommand{\section}{\secdef\sct\sect}
\newcommand{\sct}[2][default]{\refstepcounter{section}
\addcontentsline{toc}{section}
{{\tocsection {}{\thesection}{\!\!\!\!#1\dotfill}}{}}
\vspace{0.7cm}
\centerline{ 
\scshape\arabic{section}.\ #1} \nopagebreak \vspace{0.3cm}}
\newcommand{\sect}[1]{
\vspace{0.4cm} \centerline{\large\scshape\rmfamily #1}
\vspace{0.2cm}}

\renewcommand{\subsection}{\secdef\subsct\sbsect}
\newcommand{\subsct}[2][default]{\refstepcounter{subsection}
\addcontentsline{toc}{subsection}
{{\tocsection{\!\!}{\hspace{1.2em}\thesubsection}{\!\!\!\!#1\dotfill}}{}}
\nopagebreak\vspace{0.45\baselineskip} {\flushleft\bf
\arabic{section}.\arabic{subsection}~\bf #1.~}
\\*[3mm]\noindent
\nopagebreak}
\newcommand{\sbsect}[1]{\vspace{0.1cm}\noindent
\textbf{#1.~}\vspace{0.1cm}}

\renewcommand{\subsubsection}{%
\secdef \subsubsect\sbsbsect}
\newcommand{\subsubsect}[2][default]{%
\refstepcounter{subsubsection} 
\addcontentsline{toc}{subsubsection}{{\tocsection{\!\!}
{\hspace{3.05em}\thesubsubsection}{\!\!\!\!#1\dotfill}}{}}
\nopagebreak
\vspace{0.15\baselineskip} \nopagebreak {\flushleft\rmfamily
\itshape\arabic{section}.\arabic{subsection}.\arabic{subsubsection}
\ \rmfamily #1\/.}\ }
\newcommand{\sbsbsect}[1]{\vspace{0.1cm}\noindent
\rmfamily \itshape
\arabic{section}.\arabic{subsection}.\arabic{subsubsection} \
\sffamily #1\/.\ }

\iffinal
\newcommand{\printversion}{}
\else
\newcommand{\printversion}{, \version}
\fi

\newcommand{\textd}{\text{\rm d}\mkern0.5mu}
\newcommand{\texti}{\text{\rm i}\mkern0.7mu}

\renewcommand{\AA}{\mathcal A}
\newcommand{\BB}{\mathcal B}

\newcommand{\GG}{\mathcal G}

\newcommand{\CalS}{\mathcal S}

\newcommand{\E}{\mathbb E}

\newcommand{\BbbH}{\mathbb H}

\newcommand{\K}{\mathbb K}
\newcommand{\BbbL}{\mathbb L}

\newcommand{\BbbP}{\mathbb P}

\newcommand{\R}{\mathbb R}

\newcommand{\T}{\mathbb T}

\newcommand{\Z}{\mathbb Z}

\newcommand{\twoeqref}[2]{(\ref{#1}--\ref{#2})}
\newcommand{\1}{{1\mkern-4.5mu\textrm{l}}}
\renewcommand{\1}{\text{\sf 1}}

\newcommand{\scrC}{\mathscr{C}}
\newcommand{\scrF}{\mathscr{F}}
\newcommand{\scrK}{\mathscr{K}}

\newcommand{\bS}{\boldsymbol S}
\newcommand{\bsigma}{\boldsymbol\sigma}

\newcommand{\bxi}{\boldsymbol\xi}
\newcommand{\btheta}{\boldsymbol\theta}

\newcommand{\bt}{\boldsymbol t}

\newcommand{\bk}{\text{\bfseries\itshape k}\mkern1mu}
\newcommand{\br}{\text{\bfseries\itshape r}\mkern1mu}

\newcommand{\hata}{\hat{\text{\rm a}}}
\newcommand{\hatb}{\mkern-4mu\hat{\mkern4mu\text{\rm b}}}
\newcommand{\hatc}{\hat{\text{\rm c}}}
\newcommand{\hate}{\hat{\text{\rm e}}}
\newcommand{\hatw}{\hat{\text{\rm w}}}
\newcommand{\hatS}{\hat S}
\newcommand{\bhatS}{\hat\bS}

\newcommand{\tg}{\textsl{t}_{\textsl{2g}}}
\newcommand{\eg}{\textsl{e}_{\textsl{g}}}
\newcommand{\td}{\textsl{d}}

\newcommand{\cc}{{\text{\rm c}}}
\newcommand{\BBE}{\BB_{\text{\rm E}}}
\newcommand{\BBSW}{\BB_{\text{\rm SW}}}

\newcommand{\aD}{\Delta}

\begin{document}

\title[Long-range order in 120$^\circ$-model\printversion]
{\fontsize{15}{21}\selectfont Orbital ordering in transition-metal compounds:~I.~ The 120-degree model}

\author[M.~Biskup, L.~Chayes and Z.~Nussinov\printversion]
{Marek~Biskup,${}^1$\, 
Lincoln~Chayes${}^1$ \,and\, Zohar~Nussinov${}^2$}

\maketitle

\vspace{-5mm}
\centerline{${}^1$\textit{Department of Mathematics, UCLA, Los Angeles, California, USA}}
\centerline{${}^2$\textit{Theoretical Division, Los Alamos National Laboratory, Los Alamos, USA}}

\vspace{2mm}
\begin{quote}
\footnotesize {\bf Abstract:} 
We study the classical version of the 120$^\circ$-model. This is an attractive nearest-neighbor system in three dimensions with XY (rotor) spins and interaction such that only a particular projection of the spins gets coupled in each coordinate direction. 
Although the Hamiltonian has only discrete symmetries, it turns out that every constant field is a ground state.
Employing a combination of spin-wave and contour arguments we establish the existence of long-range order at low temperatures.
This suggests a mechanism for a type of ordering in certain models of transition-metal compounds where the very existence of long-range order has heretofore been a matter of some controversy.
\end{quote}
\vspace{4mm}

\vspace{-7mm}
\setcounter{tocdepth}{3}

\footnotesize

\begin{list}{}
{\setlength{\topsep}{0in}\setlength{\leftmargin}{0.34in}\setlength{\rightmargin}{0.5in}}
\item[]
\tableofcontents
\end{list}
\vspace{-1.2cm}

\begin{list}{}
{\setlength{\topsep}{0in}\setlength{\leftmargin}{0.5in}\setlength{\rightmargin}{0.5in}}
\item[]
\hskip-0.01in
\hbox to 2.3cm{\hrulefill}
\item[]
{\fontsize{8.6}{8.6}\selectfont\copyright\,\,\,2004 by M.~Biskup, L.~Chayes, Z.~Nussinov. Reproduction, by any means, of the entire article for non-commercial purposes is permitted without~charge.\vspace{2mm}}
\end{list}
\normalsize

\section{Introduction}
\vspace{-5mm}\noindent
\subsection{Overview}
For attractive classical spin systems with ground states related by an internal symmetry, ordering usually occurs by one of two mechanisms: The existence of \emph{surface tension} between thermal perturbations of the ground states or condensation of \emph{spin-wave deviations} away from the ground states. The former is most common in models where the internal symmetry is discrete, while the latter circumstances are best exhibited in systems with continuous symmetries. This paper will be concerned with an attractive spin system---the so called \emph{120$^\circ$-model}---which displays characteristics reminiscent of both phenotypes. A related model of this sort---the so called \emph{orbital compass model}---will be the subject of a continuation of this paper~\cite{Biskup-Chayes-Nussinov}.
A common feature of both systems is that the presence/absence of long-range order is all but readily apparent. 

To underscore the above (admittedly vague) allegations, let us introduce the formal Hamiltonian of the 120$^\circ$-model:
\begin{equation}
\label{Hdiff}
\mathscr{H}=\frac J2\sum_{\br}\Bigl\{\bigl(S_{\br}^{(\hata)}-S_{\br+\hate_x}^{(\hata)}\bigr)^2
+\bigl(S_{\br}^{(\hatb)}-S_{\br+\hate_y}^{(\hatb)}\bigr)^2+\bigl(S_{\br}^{(\hatc)}-S_{\br+\hate_z}^{(\hatc)}\bigr)^2\Bigr\}.
\end{equation}
Here~$\br$ is a site on the cubic lattice~$\Z^3$, the~$\bS_{\br}$'s are the usual~XY-spins, namely two-dimensional vectors of unit length, and~$\hate_x$,~$\hate_y$ and~$\hate_z$ are the lattice unit vectors in the three coordinate directions.
To define the quantities~$S_{\br}^{(\hata)}$,~$S_{\br}^{(\hatb)}$ and~$S_{\br}^{(\hatc)}$, let~$\hata$,~$\hatb$ and~$\hatc$ denote three vectors on the unit circle evenly spaced by~120$^\circ$. Then~$S_{\br}^{(\hata)}=\bS_{\br}\cdot\hata$ and similarly for $S_{\br}^{(\hatb)}$ and $S_{\br}^{(\hatc)}$.
We have $J>0$ so the interaction is ferromagnetic.

As is manifestly obvious from \eqref{Hdiff}, any \emph{constant} spin field is a ground state and since we are dealing with continuous spins, no contour-based argument readily suggests itself. (As we shall see later, there are also other ground states, but these need not concern us at the moment.) On the other hand, due to the directional bias of the coupling, a naive spin-wave argument based on the use of \emph{infrared bounds}~\cite{FSS,DLS,FILS1,FILS2} results in divergent momentum-space integrals. In particular, as we later show, the spherical version of this model has a free energy that is analytic at all temperatures.
Worse yet, the rigorous version of a disorder-by-spin-wave argument, the Mermin-Wagner theorem, requires the continuous symmetry to be present at the level of the Hamiltonian, which here is simply not the case.
Thus, the system in \eqref{Hdiff} is right on the margin.

The main goal of this paper will be to establish long-range order in this model. (Precise definitions will appear at the end of this section; precise statements of the theorems will appear in the next section.)
The mechanism for ordering involves the combination of different aspects taken from both of the classic types of arguments. Specifically, on the basis of a realistic spin-wave calculation we show that, for all intents and purposes, most of the ground states are destabilized, leaving us with only a manageable number of contenders. Among the survivors, a surface tension (with some unusual features) is established.
Thereafter, via arguments which are relatively standard, the existence of multiple states at low temperatures can be concluded. 

The described reduction of the ground state degeneracy by accounting for the free energy of the excitations is reminiscent of the problems analyzed previously in~\cite{Dinaburg-Sinai,Bricmont-Slawny,Holicky-Zahradnik}. (In our cases, the role of excitations is taken by the spin waves.) However, the model studied in this paper presents us with several novel features. For instance, unlike in~\cite{Dinaburg-Sinai,Bricmont-Slawny,Holicky-Zahradnik} which focused on discrete spin systems with ``stratified'' ground states, here we are dealing with a \emph{continuum} of homogeneous ground states related by a continuous internal symmetry. Incidentally, ``stratified'' ground states also exist in our systems, see~Sect.~\ref{sec1.3}. Here these must be ruled out on the basis of a modified spin-wave calculation which accounts for the free energy carried by deviations from inhomogeneous background.

Although the authors would have been proud to stake the claim of having concocted a model system with such an esoteric mechanism of ordering, it turns out that interest in the 120$^\circ$-model---as well as the closely related orbital compass model---is not entirely academic. Indeed, both systems arise naturally in the study of transition-metal compounds.
Here magnetic order of some type has been firmly established by experimental methods, but the nature and the mechanism for the order is unclear.
The problem persists up to the theoretical level; the question whether \emph{any} interacting model based on the physics of transition-metal orbitals is capable of supporting long-range order has heretofore been a matter of controversy.
From the present paper we now know that, at the level of finite-temperature classical spin systems, ordering indeed occurs for the 120$^\circ$-model. This strongly suggests (but of course does not prove) that a similar ordering is exhibited in the quantum and itinerant-electron versions of these models.

\vskip1.2mm
The rest of this paper is organized as follows.
In Sect.~\ref{sec1.2} we describe the physical origins of these problems.
A precise definition of the classical 120$^\circ$-model is given and the ground states are discussed in Sect.~\ref{sec1.3}.
In Sect.~\ref{sec2.1} we state our main result concerning the existence of phase transition in the 120$^\circ$-model while in Sect.~\ref{sec2.2} we outline the principal ideas of the proof of long-range order. The actual proofs are given in Sect.~\ref{sec3}. The techniques we employ are contour methods based on chessboard estimates but the infinite degeneracy of the ground states also requires us to perform some intricate spin-wave calculations. These technical details are the subject of Sects.~\ref{sec4}--\ref{sec6}. 
Sect.~\ref{sec7} collects some observations concerning the spherical version of the model at hand.

\subsection{Quantum origins}
\label{sec1.2}\noindent
In the standard description of electrons in solids, it is often the case that the accumulation of itinerant charges is heavily disfavored. This (presumably) results in localized electrons which interact only via spin exchange. In the circumstances which are most often studied, only a \emph{single} orbital is available at each site, which produces an effective antiferromagnetic interaction. However, in \emph{transition-metal compounds} (e.g., vanadates, manganites, titanates, cuprates, etc.) there are multiple essentially-degenerate orbitals any of which could be occupied. In particular, if the transition metal ion interacts with a local environment which is of octahedral symmetry, the~$3\td$-quintet of the transition-metal ion is split into a low-lying triplet---the~$\tg$ orbitals---and a pair---the~$\eg$ orbitals---of considerably higher energy. 

In the absence of any other significant effects, one circumstance which is amenable to further approximation is when there is but a single electronic degree of freedom per site. The two obvious distinguished cases are the $\eg$ and~$\tg$ compounds. The former will come about under two conditions: First, if the~$\tg$ orbitals are filled and there is one extra electron per site to occupy the~$\eg$ orbitals. Second, same as above but here there are three electrons (out of possible total of four) in the~$\eg$ orbitals and the role of the single electronic degree of freedom is played by the hole. The latter cases, the~$\tg$ compounds, occur if the~$\eg$ orbitals are empty and there is a single electron or a hole in the~$\tg$ orbitals. It appears that the situations leading to~$\eg$-type compounds are far more prevalent.

In any of the above circumstances, one can write down the inevitable itinerant electron model describing the spins and orbitals. After the standard superexchange calculation/approximation---analogous to that which relates the single-orbital models to the Heisenberg antiferromagnets---we arrive at a problem which involves ``only'' quantum spins. Of course, in these models there will be two types of quantum states. Namely, those corresponding to the actual (electronic) spin degrees of freedom and those corresponding to the occupation numbers of the dynamical orbitals. The resulting system is described by the Kugel-Khomskii~\cite{Kugel-Khomskii} Hamiltonian
\begin{equation}
\label{qHam}
\mathscr{H} = J \sum_\alpha\sum_{\br} \bigl(\hat\pi_{\br}^\alpha \hat\pi_{\br+\hate_\alpha}^\alpha - 
\tfrac12\hat\pi_{\br}^\alpha - \tfrac12\hat\pi_{\br+\hate_\alpha}^\alpha
+ \tfrac14\bigr) \bigl(\bsigma_{\br} 
\cdot \bsigma_{\br+\hate_\alpha} + 1\bigr).
\end{equation}
Here the interaction takes place at the neighboring sites of the cubic lattice~$\Z^3$ representing the positions of the transition-metal ions, the object~$\bsigma_{\br}$ is the triple of the usual Pauli matrices acting on the spin degrees of freedom at the site~$\br$ and the~$\hat\pi_{\br}^\alpha$ are pseudospin operators acting on the orbital degrees of freedom at the site~$\br$.
As usual, the vectors~$\hate_x$,~$\hate_y$ and~$\hate_z$ are the unit vectors in the principal lattice directions.

Depending on which of the orbitals play the seminal role, the two choices for the orbital pseudospins are
\begin{equation}
\label{1.2}
\hat\pi_{\br}^\alpha = 
\begin{cases}
\tfrac14 (- \sigma^z \pm \sqrt{3} \sigma^x),\qquad&\text{if }\alpha=x,y,
\\*[1mm]
\tfrac12 \sigma^z,\qquad&\text{if }\alpha=z,
\end{cases}
\end{equation}
for the~$\eg$-compounds, while 
\begin{equation}
\label{1.3}
\hat\pi_{\br}^\alpha=\tfrac12\sigma^\alpha,\qquad\text{for}\quad \alpha=x,y,z,
\end{equation}
for the~$\tg$-compounds.
The former choice gives rise to the \emph{120$^\circ$-model} and the latter to the \emph{orbital compass model}.

\smallskip
The question of obvious importance is to prove/disprove the case for ordering of the spins or orbitals in these models. In this vain, it should be remarked that the orbital-compass version of the Kugel-Khomskii Hamiltonian---if reformulated back in the language of itinerant-electron model---has some unapparent symmetries. For instance, as pointed out by Harris \textit{et al} \cite{Harris}, the total spin of electrons in $\alpha$-orbitals at sites of any plane orthogonal to the direction represented by~$\alpha$ is a conserved quantity.
On the basis of these symmetries, a Mermin-Wagner argument has been used \cite{Harris} to show that, in the three-dimensional system, the spin variables represented by~$\bsigma_{\br}$ in \eqref{qHam} \emph{cannot} order. 

Notwithstanding the appeal of this ``no-go'' result, we note that the absence of spin order does not preclude the more interesting possibility of orbital ordering in these systems. Indeed, on the experimental/theoretical front, it appears that there is a reasonable consensus ``for'' orbital ordering; the references \cite{Caciuffo02,Castellani78,Pen97,vdBrink99,Ishihara98,Elfimov99,Okamoto02} constitute works which support this picture while the references \cite{Keimer00,Larochelle01,Benfatto99,Takahashi03} offer arguments that dispute or down-play the role of orbital order in the magnetic properties of transition-metal compounds. We refer to (slightly biased) review articles~\cite{Tokura-Nagaosa, vdBrink-review} for more information.

\smallskip
In order to study the phenomenon of orbital ordering in the context of Kugel-Khomskii models, the interactions are often further reduced. Neglecting all sorts of terms including all terms pertaining to intrinsic spin, the resulting \emph{orbital-only} model has the Hamiltonian
\begin{equation}
\label{oHam}
\mathscr{H} = J \sum_\alpha\sum_{\br} \hat\pi_{\br}^\alpha \hat\pi_{\br+\hate_\alpha}^\alpha.
\end{equation}
Here, as before, $\hat\pi_{\br}^\alpha$ are as in \eqref{1.2} for the 120$^\circ$-model and \eqref{1.3} for the orbital compass model. Full physical justification of these approximations goes beyond the scope of this paper.

Interestingly enough, the Hamiltonian \eqref{oHam} for the~120$^\circ$-case can be arrived at by entirely different means. In particular, among the other ``competing'' mechanisms so far omitted from the discussion is the \emph{Jahn-Teller effect} which refers to further distortion of octahedral geometry of the ``crystal field'' surrounding the transition-metal ions. On the basis of symmetry considerations it has been argued~\cite{Khomskii-Mostovoy}
that, in the~$\eg$-compounds, this will lead to an effective interaction among the nearby orbitals which turns out to be exactly of the type \eqref{oHam}. In the rare cases of the~$\tg$-compounds \emph{with} Jahn-Teller effects, it turns out that yet another Hamiltonian emerges. In the $\tg$-cases the interplay of the two interactions must be properly accounted for; in contrast to the~$\eg$-situations where, no matter what, we get the 120$^\circ$-model. For these and other reasons---the latter mostly concerning the ``degree'' of difficulty---the remainder of this paper will be focused on the~120$^\circ$-model.

\subsection{The classical models}
\label{sec1.3}\noindent
The classical versions of the above orbital models can be obtained from their quantum counterparts by replacing the operators~$\hat\pi_{\br}^\alpha$ by appropriate projections of the classical spin variables~$\bS_{\br}$, which live on the unit sphere in~$\R^n$. A standard justification for the classical approximation is via the ``$S\to\infty$'' limit; cf~\cite{Lieb,Simon2,Duffield} and also \cite{Dyson,Bruno1,Bruno2} for some results in this direction. As was the case for the quantum systems, there are two major types of models under consideration: Classical 120$^\circ$-model and classical orbital compass model. We proceed with formal definitions.

\smallskip
Let $\Z^3$ denote the three-dimensional cubic lattice and let $\bS_{\br}$, where $\br\in\Z^3$, be unit vectors in~$\R^2$.
We let $\hata$, $\hatb$ and $\hatc$ denote three evenly-spaced vectors on the unit circle, for instance,
\begin{equation}
\label{abc}
\hata=(1,0),\qquad
\hatb=\bigl(-\tfrac12,\tfrac{\sqrt3}2\bigr)
\quad\text{and}\quad
\hatc=\bigl(-\tfrac12,-\tfrac{\sqrt3}2\bigr),
\end{equation}
and define the projections $S_{\br}^{(\hata)}=\bS_{\br}\cdot\hata$, where the dot denotes the usual dot product, and similarly for  $S_{\br}^{(\hatb)}$ and  $S_{\br}^{(\hatc)}$.
In this notation, the (formal) Hamiltonian of the 120$^\circ$-model is given by
\begin{equation}
\label{H120}
\mathscr{H}=-J\sum_{\br}\bigl(S_{\br}^{(\hata)}S_{\br+\hate_x}^{(\hata)}
+S_{\br}^{(\hatb)}S_{\br+\hate_y}^{(\hatb)}+S_{\br}^{(\hatc)}S_{\br+\hate_z}^{(\hatc)}\bigr),
\end{equation}
with again $J>0$.
For convenience we will sometimes label the lattice direction and the spin direction with the same index; i.e.,~$S_{\br}^{(\alpha)}$, $\alpha=1,2,3$, meaning, e.g.,~$S_{\br}^{(\hatb)}$ for~$\alpha=2$, etc. Then \eqref{H120} can be written
\begin{equation}
\label{HOC}
\mathscr{H}=-J\sum_{\br,\alpha}S_{\br}^{(\alpha)}S_{\br+\hate_\alpha}^{(\alpha)}.
\end{equation}
We remark in passing that \eqref{HOC} is also the form of the orbital compass Hamiltonian, only in this case the~$\bS_{\br}$, $\br\in\Z^3$, are genuine (three-component) Heisenberg spins and the upper index of the spin stands for its \emph{Cartesian} component.

\begin{remark}
The 120$^\circ$-model (as well as its orbital compass counterpart) can be generalized to hypercubic lattices in other dimensions as well as to other graphs. For instance, in four spatial dimensions the spins are from the unit sphere in~$\R^3$ and the interaction in the various lattice directions is via the projections of the spins onto the vectors pointing from the origin to the vertices of an appropriately centered tetrahedron. However, these variant situations are fairly difficult geometrically and since they do not always correspond to the structure of the original quantum-spin model, we will not consider them in this paper.
\end{remark}

\newcounter{obrazek}

\begin{figure}[t]
\refstepcounter{obrazek}
\label{fig1}
\vspace{.2in}
\ifpdf
\centerline{\includegraphics[width=3.4in]{groundstates.pdf}}
\else
\centerline{\includegraphics[width=3.4in]{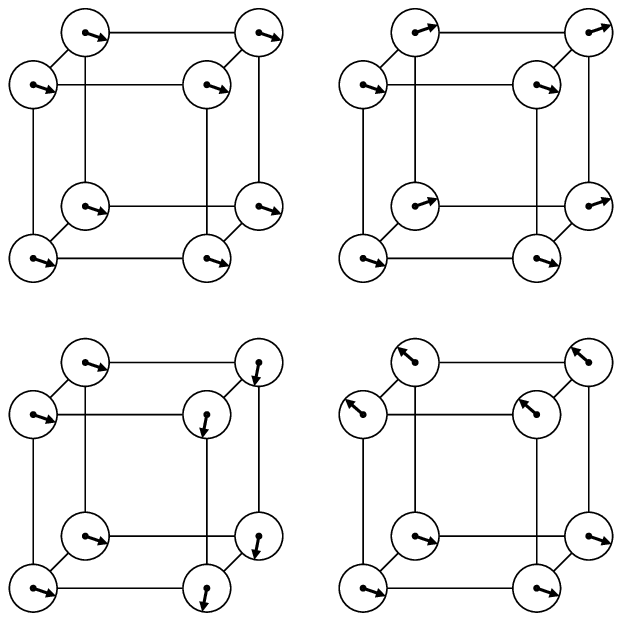}}
\fi
\vspace{.1in}
\bigskip
\begin{quote}
\fontsize{9}{5}\selectfont
{\sc Figure~\theobrazek.\ }
A picture demonstrating the discrete symmetries of the ground states in the 120$^\circ$-model on a cube with one spin fixed. Here the horizontal and vertical directions correspond to the~$y$ and~$z$-axes, respectively; the front face of the cubes is perpendicular to the~$x$-axis. The upper-left cube is simply the homogeneous ground state, the upper-right cube has a spin reflection in the~$\hata$-direction as one moves in the $\hate_x$-direction across the $yz$-midplane. The bottom cubes have analogous~$\hatb$ and~$\hatc$-reflections. The structure of any (global) ground state is demonstrated by checking for consistency between all neighboring cubes.
\normalsize
\end{quote}
\end{figure}

The salient feature of the 120$^\circ$-model (as well as the orbital compass model) is that the ground-state space of the Hamiltonian is \emph{infinitely degenerate}.
This is manifest if we write the Hamiltonian in the form \eqref{Hdiff} which follows immediately from \eqref{H120} by the fact that for any $\bS_{\br}$ from the unit circle in~$\R^2$,
\begin{equation}
[S_{\br}^{(\hata)}]^2+[S_{\br}^{(\hatb)}]^2+[S_{\br}^{(\hatc)}]^2=\frac32.
\end{equation}
It is now apparent that any constant vector field receives the minimum possible energy---namely zero---from the Hamiltonian in \eqref{Hdiff}. 

Unfortunately, as we remarked before, the ground state situation is further complicated by the fact that the constant configurations are certainly not the only minimum-energy states available in this system. For instance, it is easy to verify that, starting from a constant configuration, the reflection of all spins in an $xy$-plane ``through'' vector~$\hatc$ preserves the overall energy. (Here, the $\hatc$-projection is not affected by this procedure and the~$\hata$ and~$\hatb$-projections just swap their roles.) Hence, plenty of other ground states can be obtained from the constant ones by reflecting all spins in a collection of parallel lattice planes; see Fig.~\ref{fig1} for some examples.
Notwithstanding, as will be proved later, these non-translation invariant ground states are disfavored by the onslaught of positive temperatures.

\begin{remark}
The ground state situation is yet more intricate in the orbital compass model which is foremost among the reasons that our analysis of this system was postponed.
\end{remark}

\subsection{Gibbs measures}
The (still formal) Hamiltonian in \eqref{Hdiff} can be used to define the Gibbs measures for the corresponding spin system. Explicitly, let $\Lambda\subset\Z^3$ be a finite set and let $\partial\Lambda$ denote the set of sites in $\Z^3\setminus\Lambda$ that have an edge with one endpoint in~$\Lambda$. Given a spin configuration $\bS_\Lambda$ in~$\Lambda$ and a boundary condition $\bS_{\partial\Lambda}$ on~$\partial\Lambda$, we let $\mathscr{H}_\Lambda(\bS_\Lambda|\bS_{\partial\Lambda})$ be the restriction of the sum in \eqref{Hdiff} to~$\br$ and~$\alpha$ such that $\br\in\Lambda$ or $\br+\hate_\alpha\in\Lambda$ (or both). Then the finite-volume Gibbs measure in~$\Lambda$ with boundary condition~$\bS_{\partial\Lambda}$ is a measure on the configurations $\bS_\Lambda=(\bS_{\br})_{\br\in\Lambda}$ given by
\begin{equation}
\label{1.11}
\mu_\Lambda^{(\bS_{\partial\Lambda})}(\textd\bS_\Lambda)=
\frac{e^{-\beta\mathscr{H}_\Lambda(\bS_\Lambda|\bS_{\partial\Lambda})}}
{Z_{\Lambda,\beta}^{\bS_{\partial\Lambda}}}
\prod_{\br\in\Lambda}\Omega(\textd\bS_{\br}).
\end{equation}
Here $\Omega$ is the Lebesgue measure on the unit circle (in other words, $\Omega$ is the \emph{a priori} spin distribution) and $Z_{\Lambda,\beta}^{\bS_{\partial\Lambda}}$ is the corresponding normalization constant.

Regarding these measures as the so-called specifications, the DLR-formalism can be used to define the infinite-volume Gibbs measures (aka Gibbs states). Explicitly, the latter are probability measures on configurations $\bS_{\br}$ for $\br\in\Z^d$, whose conditional probability in a finite volume given a boundary condition $\bS_{\partial\Lambda}$ is the measure \eqref{1.11}, for almost every~$\bS_{\partial\Lambda}$. We refer to~\cite{Georgii} for a comprehensive treatment of these concepts.
To adhere with mathematical-physics terminology, we will denote expectation with respect to the infinite-volume Gibbs measures by~$\langle-\rangle$.

\section{Main results}
\label{sec2}\noindent
Here we state the main theorem of this paper and provide heuristic reasoning for the existence of long-range order in the system at hand. We also provide some more discussion and remarks on literature concerning the related problems that have previously been studied.

\subsection{Long-range order in the 120-degree model}
\label{sec2.1}\noindent
Let $\hatw_\tau$, $\tau=1,2,\dots,6$, denote the six vectors on the unit circle in~$\R^2$ corresponding to the six sixth-roots of unity. Explicitly, we define
\begin{equation}
\label{2.1}
\hatw_\tau=\bigl(\cos(\tfrac\pi3\tau),\sin(\tfrac\pi3\tau)\bigr),
\qquad \tau=1,2,\dots,6.
\end{equation}
The principal result of this paper is then as follows:

\begin{theorem}
\label{thm2.1}
Consider the 120$^\circ$-model with a fixed coupling constant~$J>0$. Then there exits a number $\beta_0\in(0,\infty)$ and a function $\beta\mapsto\epsilon(\beta)\in[0,1)$ with $\epsilon(\beta)\to0$ as $\beta\to\infty$ such that the following is true: For all $\beta\ge\beta_0$ there exist six distinct, infinite-volume, translation-invariant Gibbs states $\langle-\rangle_{\beta,J}^\tau$, with $\tau=1,2,\dots,6$, such that
\begin{equation}
\label{lbexp}
\langle \hatw_\tau\cdot\bS_{\br}\rangle_{\beta,J}^\tau\ge 1-\epsilon(\beta),
\qquad \tau=1,2,\dots,6,
\end{equation}
is valid for all $\br\in\Z^3$.
\end{theorem}

We note that once $\langle \hatw_\tau\cdot\bS_{\br}\rangle_{\beta,J}^\tau\ne0$, we must have that $\langle \bS_{\br}\rangle_{\beta,J}^\tau\ne\boldsymbol0$. Consequently, \eqref{lbexp} implies the existence of a long-range order because at $\beta\ll1$, the standard high-temperature expansions (or Dobrushin uniqueness techniques, see~\cite[Theorem V.1.3]{Simon}), imply that $\langle \bS_{\br}\rangle_{\beta,J}=\boldsymbol0$ in any Gibbs state~$\langle-\rangle_{\beta,J}$. Moreover, as $\beta\to\infty$, the measure corresponding to $\langle-\rangle_{\beta,J}^\tau$ gets increasingly concentrated around~$\hatw_\tau$.

\smallskip
Theorem~\ref{thm2.1} is proved in Sect.~\ref{sec3.3} subject to some technical claims whose proof is postponed to Sect.~\ref{sec6.2}.

\subsection{Spin-wave heuristics}
\label{sec2.2}\noindent
Here we provide a heuristic outline of the spin-wave reasoning which ultimately leads to the proof of the above theorem. The precise version of the argument is given in Sects.~\ref{sec4} and~\ref{sec5}.

\smallskip
The starting point of our analysis differs in perspective from the usual sort of spin-wave arguments which have previously been the subject of mathematical theorems. In the standard approaches, one attempts to rewrite the \emph{full} Hamiltonian as a ``spin-wave'' Hamiltonian, carry out a calculation and control the errors later (if at all). An extreme example of this is the \emph{spherical model} whose working definition is ``the spin system for which the spin-wave approximation is exact.'' However, as alluded to previously, this sort of spin-wave approximation is inadequate to capture the essential features of the problem at hand. (See Sect.~\ref{sec7} for more details.)

The present perspective, which is standard in condensed matter physics but has not yet been the subject of detailed mathematical analysis, can be summarized as follows: We will collect the important excitations about the various ground states into spin-wave modes. These modes form the basis of an approximate low-temperature expansion which via the standard arguments yields the existence of several low-temperature states.

\smallskip
Let us start by expressing all spins in terms of angular variables, i.e., $\bS_{\br}=(\cos\theta_{\br},\sin\theta_{\br})$. The (homogeneous) ground states are then just $\theta_{\br}=\text{const}=\theta^\star$. We define $\xi_{\br}=\theta_{\br}-\theta^\star$ so that, in the~$x$~direction, the interaction is given by $\frac12(\beta J)[\cos(\theta^\star+\xi_{\br})-\cos(\theta^\star+\xi_{\br+\hate_x})]^2$ with analogous formulas in the $y$ and~$z$ directions.
Thus, to leading order in~$\xi_{\br}$'s, we have
\begin{equation}
\frac{\beta J}2\bigl(S_{\br}^{(\hata)}-S_{\br+\hate_x}^{(\hata)}\bigr)^2
\approx\frac{\beta J}2\sin^2(\theta^\star)(\xi_{\br}-\xi_{\br+\hate_x})^2
\end{equation}
and, similarly,
\begin{equation}
\begin{aligned}
\frac{\beta J}2\bigl(S_{\br}^{(\hatb)}-S_{\br+\hate_y}^{(\hatb)}\bigr)^2
&\approx\frac{\beta J}2\sin^2(120^\circ-\theta^\star)(\xi_{\br}-\xi_{\br+\hate_y})^2,
\\*[2mm]
\frac{\beta J}2\bigl(S_{\br}^{(\hatc)}-S_{\br+\hate_z}^{(\hatc)}\bigr)^2
&\approx\frac{\beta J}2\sin^2(120^\circ+\theta^\star)(\xi_{\br}-\xi_{\br+\hate_z})^2.
\end{aligned}
\end{equation}

We will encode the $\theta^\star$-dependence into effective coupling parameters:
$q_1=\sin^2(\theta^\star)$,~$q_2=\sin^2(\theta^\star-120^\circ)$ and~$q_3=\sin^2(\theta^\star+120^\circ)$.
Then the effective interaction for deviations about the~$\theta^\star$-state can be written as
\begin{equation}
\beta\mathscr{H}_{\text{eff}}^{(\theta^\star)}(\bxi)=\frac{\beta J}2\sum_\alpha\sum_{\br}q_\alpha(\xi_{\br}-\xi_{\br+\hate_\alpha})^2.
\end{equation}
Therefore, in some approximate sense, the partition function for deviations about the state where the spins are pointing in the direction~$\theta^\star$ can be written as
\begin{equation}
\label{ZLheur}
Z_{L,\beta}(\theta^\star)\approx\int \textd\bxi\, e^{-\beta\mathscr{H}_{\text{eff}}^{(\theta^\star)}(\bxi)}
\end{equation}
where $\textd\bxi$ denotes the product Lebesgue measure.

As we will see, the integral is, as it stands, somewhat ill defined because the Hamiltonian provides no decay for the zero Fourier mode of~$\bxi$.
However, it is recalled that for the above derivation to be meaningful, the~$\xi_{\br}$'s had to be fairly small. So, one way out---which is what we will do in our proofs---is to restrict the integration measure in \eqref{ZLheur} only to (the Cartesian product of) small intervals centered at zero. Another way out, which leads to more transparent calculations, is to define the full object $Z_{L,\beta}(\theta^\star)$ as the partition function \emph{constrained} to configurations where, say, the average spin \emph{equals} $(\cos\theta^\star,\sin\theta^\star)$. (As we will see, inserting the appropriate $\delta$-function on the right-hand side of \eqref{ZLheur} permits us to integrate the~$\xi_{\br}$'s over all real values.) In this language, the said constraint reads $\sum_{\br}\xi_{\br}=0$, i.e., no ``zero mode.'' For future reference, we denote the right-hand side of \eqref{ZLheur} with this constraint enforced by
\begin{equation}
\Bigl(\frac{2\pi}{\beta J}\Bigr)^{L^3/2}e^{-L^3 F_L(\theta^\star)}.
\end{equation}
The reason for the prefactor will become clear momentarily.

The translation-invariant structure of the effective Hamiltonian (and the constraint) prompts us to use the Fourier-transformed variables,
\begin{equation}
\widehat\xi_{\bk}=\frac1{L^{3/2}}\sum_{\br}\xi_{\br}\,e^{\texti\bk\cdot\br},
\end{equation}
where~$\bk$ takes values in~$\T_L^\star=\{2\pi L^{-1}(n_1,n_2,n_3)\colon -L/2\le n_1,n_2,n_3\le L/2\}$ which is known as the reciprocal volume (or the Brillouin zone).
In terms of these variables, and the various other quantities defined, an appropriate spin-wave Hamiltonian can be constructed:
\begin{equation}
\beta\mathscr{H}_{\text{\rm SW}}(\widehat\bxi)=\frac{\beta J}2
\sum_{\bk\in\T_L^\star\smallsetminus\{\boldsymbol 0\}}|\widehat\xi_{\bk}|^2
\biggl(\sum_\alpha q_\alpha|1-e^{\texti k_\alpha}|^2\biggr),
\end{equation}
where we have made explicit the absence of the contribution from the ``zero mode.''

The calculation is now standard and we get
\begin{equation}
\Bigl(\frac{2\pi}{\beta J}\Bigr)^{L^3/2}e^{-L^3 F_L(\theta^\star)}
=\prod_{\bk\in\T_L^\star\smallsetminus\{\boldsymbol 0\}}
\Bigl[\frac{2\pi}{\beta J\sum_\alpha q_\alpha|1-e^{\texti k_\alpha}|^2}\Bigr]^{1/2}.
\end{equation}
Thus, taking logs and letting $L\to\infty$, we arrive at the limiting version of~$F_L$,
\begin{equation}
\label{Fn1}
F(\theta^\star)=\frac12\int_{[-\pi,\pi]^3}
\frac{\textd\bk}{(2\pi)^3}\log\Bigl\{\sum_\alpha q_\alpha|1-e^{\texti k_\alpha}|^2\Bigr\}.
\end{equation}
This is the \emph{spin-wave free energy} for fluctuations about the direction~$\theta^\star$.

\smallskip
It is apparent that~$\theta^\star\mapsto F(\theta^\star)$ is invariant under the shift~$\theta^\star\to\theta^\star+60^\circ$. Far less obvious (but nevertheless true) is the fact that the absolute minima of~$F$ occur at $\theta^\star=0^\circ, 60^\circ,120^\circ,\dots, 300^\circ$. Thus we must conclude that, when finite temperature effects are accounted for, six ground states are better off than any of the others.
Sects.~\ref{sec4}--\ref{sec5} will be devoted to a rigorous proof of this heuristic. A similar calculation allows us to estimate the spin-wave free energy for the inhomogeneous ground states and show that these are always less favorable than the homogeneous~ones.

Notwithstanding the appeal of the spin-wave heuristic, the above is just one step of the proof.
In order to make use of spin-wave calculations, we resort to some (rather standard) contour estimates. Informally, we partition the ``world'' (by which we mean the torus) into blocks and mark those blocks where the spin configuration either features too much energy or has the characteristics of an environment without enough entropy. By adjusting the block scale we can make the penalty for marked blocks sufficient to carry out a Peierls argument. The principal tool for decoupling the correlations between various boxes is provided by the chessboard estimates (which allow, via Cauchy-Schwarz type inequalities, to estimate the probabilities of various block events by their associated constrained partition functions). Explicit details are to be found in Sect.~\ref{sec6}.

\begin{remark}
It is noted that if the reader is willing to preaccept the forthcoming treatment as fact, an interesting feature concerning the \emph{surface tension} is bound to arise. Indeed, let us imagine that the system is forced, e.g., via boundary conditions, to exhibit two favored states in the same vessel. The price for these circumstances will be the region---the interfacial region---where spins are bad. If~$\beta\gg1$, the energetic form of ``badness'' can be ruled out \emph{a fortiori}, but now we emphasize that the free energy \emph{difference} between the most and least favored states is independent of the temperature indicating that the cost of the interface will be temperature independent. Hence we anticipate that the stiffness (and also the correlation length) stays uniformly bounded away from zero and infinity as~$\beta\to\infty$.
\end{remark}

\subsection{Discussion}
\label{sec2.3}\noindent
The model under consideration exhibits infinitely many ground states, a problem which for mathematical physics has surfaced but a few times in the past. When these situations arise, the finite-temperature fate of each ground state is typically decided by its capacity to harbor excitations. Here, the dominant excitations are exactly the spin waves from the last section---the spin-wave calculation shows that only a finite number from the initial continuum of ground states survive at positive temperatures. Unfortunately, an extra complication arises due to the inhomogeneous ground states discussed in Sect.~\ref{sec1.3}. Here chessboard estimates allow us to \emph{bound} the relevant spin-wave contribution by the spin-wave free energy against a periodic background. We remark that there are systems for which the spin-wave analysis featured herein may be performed without the complication of inhomogeneous ground states. One such example is the subject of the forthcoming paper~\cite{Biskup-Chayes-Kivelson}.

As already noted, the ``entropic-selection'' mechanisms for long-range order are not new. Indeed, there have been some previous studies of the~ANNNI models and other systems exhibiting infinite degeneracy of the ground state~\cite{Dinaburg-Sinai,Bricmont-Slawny,Holicky-Zahradnik}. However, the techniques involved in~\cite{Dinaburg-Sinai,Bricmont-Slawny,Holicky-Zahradnik} are based on the premise that there is a substantial gap in the energy spectrum which separates the excitations resolving the ground state degeneracy from the remaining ones. Due to the continuous nature of the spins, and the symmetry of the ground states, no such gap is of course present for the 120$^\circ$-model. Instead, a decisive contribution to the entropic content comes from long wave-length excitations, i.e., the aforementioned spin waves.

Another set of problems which are related to the present paper are the models with continuous spins studied in~\cite{Dobrushin-Zahradnik,Zahradnik}. There the spins are \emph{a priori} Gaussian random variables with covariance given by the inverse lattice Laplacian and with an on-site (anharmonic) potential. However, this potential is required to have only a finite number of nearly-quadratic minima (all of which have a uniformly positive curvature) which necessarily implies only a finite number of low temperature states. Notwithstanding, the work in~\cite{Dobrushin-Zahradnik,Zahradnik} exemplifies situations where a ground state degeneracy is lifted by spin-wave-like excitations resulting in a reduced number of Gibbs states at positive temperatures. It is quite possible that the Pirogov-Sinai techniques used in \cite{Dinaburg-Sinai,Bricmont-Slawny,Holicky-Zahradnik,Dobrushin-Zahradnik,Zahradnik} can after some work be adapted to our cases. However, at present the arguments via chessboard estimates seem considerably easier.

As noted in Sect.~\ref{sec1.2}, the motivation to study these systems comes from the observed magnetic behavior of transition metal compounds. A complete understanding of these systems may therefore require a full quantum-mechanical treatment. We expect a similar mechanism for ordering to be present also in the quantum-mechanical version of the 120$^\circ$-model (as well as the orbital compass model). However, the only method of proof that seems promising in this context is the Pirogov-Sinai expansion of some sort. A general theory of these expansions for quantum systems exists, both for the situations with~\cite{DFF2,KU} or without~\cite{DFF1,BKU} infinite degeneracy of the ground state. But, as is the case for the classical systems, some fairly non-trivial generalizations of the existing tools would probably be necessary.

\section{Proofs of main results}
\label{sec3}\noindent
In this section we will give the proof of our main theorem, subject to some technical results which will be proved later. In particular, in Sect.~\ref{sec3.2} we define the notion of a ``bad'' box and state without a proof the principal bound concerning the simultaneous occurrence of several bad boxes; see Theorem~\ref{thm3.1}. This will be sufficient material for the proof of Theorem~\ref{thm2.1}. The proof of Theorem~\ref{thm3.1} is the subject of Sects.~\ref{sec4}-\ref{sec6}; the actual proof comes in Sect.~\ref{sec6.2}. 

\subsection{Good and bad events}
\label{sec3.2}\noindent
Here we will provide some mathematical foundations for our notions of the stable states and the contours that separate them. We will need three different scales---two of them spin-deviation scales and one a scale for the blocks which will be the setting of our various events. 

\smallskip
We will start with the fundamental spin-deviation scale which we denote by~$\Gamma$. Here we are seeking a~$\Gamma$ (which is small) such that if \emph{all} neighboring pairs of spins are within a distance~$\Gamma$ of each other, the harmonic approximation is ``good'' while if a neighboring pair violates this condition the energetic cost is drastic. On the basis of naive Taylor expansions---which is ultimately all we will do---it is clear that the latter is achieved if $\beta\Gamma^2\gg1$ and the former if $\beta\Gamma^3\ll1$. Thus, of course, we need~$\beta$ to be large and we can envision~$\Gamma$ to scale as any inverse power of~$\beta$ between~$1/3$~and~$1/2$. 

The second deviation scale will be denoted by~$\kappa$ and will serve to define sets of configurations which are effectively in one of the stable ground states. The third scale is the number~$B$ which will be used to define the spatial size of our block events. For fixed~$\kappa$, it appears that the only necessary requirement is that $\beta\Gamma^2\gg\log B$ which will always hold eventually. Unfortunately, there is some spurious interplay between the parameters~$B$, $\kappa$ and~$\Gamma$ which could, in principle, be removed in a more refined analysis. But, for this work, we will keep the ``smallness'' of~$\kappa$ in the realm of the existential and require~$B$ to get large, but only very slowly, as~$\beta$ goes to infinity.

\smallskip
In order to make our main technique, the \emph{chessboard estimates}, available we have to confine ourselves to systems with periodic boundary conditions. Let thus~$\T_L$ denote the three-dimensional torus of scale~$L$, i.e., $\T_L=\Z^3/(L\Z)^3$. In general, we will be dealing with certain events taking place in blocks of a specific scale~$B$ and we will be using the chessboard estimates to bound probabilities of these events. These blocks will be translates of the block $\Lambda_B\subset\T_L$ which we define as the cube of $(B+1)^3$ sites with the ``lowest left-most'' site at the origin.
It will be convenient (although presumably not strictly necessary) to assume that the linear scale of our finite-volume system,~$L$, when divided by~$B$ results in a power of two.

\smallskip
Now we are ready to state the definition of a ``good'' block:

\begin{definition}
\label{def1}
Let~$B$ denote a positive integer and let $\kappa>0$ and $\Gamma>0$ be sufficiently small. We will say that the spin configuration in the block~$\Lambda_B$ (or the block itself) is \emph{good} if the following two conditions are met:
\settowidth{\leftmargini}{(11)}
\begin{enumerate}
\item[(a)]
For each~$\alpha\in\{1,2,3\}$ and any neighboring pair $\br$ and $\br+\hate_\alpha$ in~$\Lambda_B$,
\begin{equation}
\label{3.1}
|S^{(\alpha)}_{\br}-S^{(\alpha)}_{\br+\hate_\alpha}|<\Gamma.
\end{equation}
\item[(b)]
All spins in~$\Lambda_B$ make an angle which is less than~$2\kappa$ from one of the preferred six directions. Explicitly, if~$\bS_{\br}=(\cos\theta_{\br},\sin\theta_{\br})$ then, for some~$\tau=1,\dots,6$, we have~$|\theta_{\br}-\tfrac{2\pi}3\tau|<2\kappa$ for all~$\br\in\Lambda_B$. Here, of course,~$\theta_{\br}$ is only determined modulo~$2\pi$.
\end{enumerate}
\end{definition}

We denote by~$\GG=\GG_{B,\kappa,\Gamma}$ the event that the block~$\Lambda_B$ is good. The complementary event, marking the situation when the block is \emph{bad}, will be denoted by inevitable~$\BB$. Our goal will be to bound various probabilities involving bad events. The main tool for these bounds will be the chessboard estimates whose basic setup and principal result we will now describe.

Let~$\BbbP_{L,\beta}$ and~$\langle-\rangle_{L,\beta}$ denote the (Gibbs) probability measure, respectively, the corresponding expectation according to the Hamiltonian \eqref{Hdiff} at inverse temperature~$\beta$ on~$\T_L$.
Let~$\bt$ denote a vector with integer coefficients identified modulo~$L/B$---in formal notation, $\bt\in\T_{L/B}$---and let~$\BB$ be an event discussed above. Then we let $\vartheta_{\bt}(\BB)$ denote the event~$\BB$ translated by the vector~$B\bt$. (For general events~$\AA$ defined on the configurations in~$\Lambda_B$ we will need an enhanced definition of~$\vartheta_{\bt}(\AA)$; cf the definition prior to Theorem~\ref{thm6.1}.) Note that if $\vartheta_{\bt}(\BB)$ and $\vartheta_{\bt'}(\BB)$ are ``neighboring'' translates of~$\BB$, then these two events both depend on the spin configuration on the shared face of the corresponding translates of~$\Lambda_B$.

\smallskip
The principal result of this section, which is the starting point for all subsequent results of this work, is the following theorem:

\begin{theorem}
\label{thm3.1}
Consider the 120$^\circ$-model as defined by \eqref{1.11}. For each sufficiently small~$\kappa>0$ and each $\eta\in(0,1)$ there exist~$L_0\in(0,\infty)$ and $\beta_0\in(0,\infty)$ and, for any any $\beta\ge\beta_0$, there exist numbers $\Gamma\in(0,1)$ and $B\in(0,\infty)$ such that the following holds: If~$\BB$ is the event---defined using~$\kappa$,~$\Gamma$ and~$B$---that the configuration in~$\Lambda_B$ is bad and $\bt_1,\dots,\bt_m$ are distinct vectors from~$\T_{L/B}$, then for any~$L\ge L_0$,
\begin{equation}
\label{3.4aa}
\BbbP_{L,\beta}\bigl(\vartheta_{\bt_1}(\BB)\cap\dots\cap\vartheta_{\bt_m}(\BB)\bigr)\le\eta^m.
\end{equation}
\end{theorem}

This result provides a way to estimate the probability of simultaneous occurrence of several bad events.
The non-trivial part of the proof of Theorem~\ref{thm3.1} boils down to the spin-wave calculations outlined in Sect.~\ref{sec2.2}. The rigorous version of these calculations requires some substantive estimations and the actual proof is therefore deferred to Sect.~\ref{sec6.2}.

\subsection{Proof of long-range order}
\label{sec3.3}\noindent
Now we are ready to prove our main theorems. We note that there are six disjoint ways to exhibit a good block for the 120$^\circ$-model, each corresponding to one of the vectors~$\hatw_\tau$. We will denote the corresponding events by~$\GG_\tau$, with $\tau=1,2,\dots,6$. Explicitly,
\begin{equation}
\GG_\tau=\GG\cap\bigl\{\bS\colon\bS_r\cdot\hatw_\tau>\cos(2\kappa),\,\br\in\Lambda_B\bigr\}.
\end{equation}
The core of the proof is the following (almost direct) consequence of Theorem~\ref{thm3.1}:

\begin{lemma}
\label{lemma3.2}
Consider the 120$^\circ$-model on~$\T_L$ and suppose that $\kappa\ll1$.
There exists a function $h\colon[0,1)\to[0,\infty)$ satisfying $h(\eta)\to0$ as $\eta\downarrow0$, such that for each sufficiently small~$\eta>0$ and each~$\beta$,~$\Gamma$ and~$B$ as allowed by Theorem~\ref{thm3.1} the following is true:
For any $\bt_1,\bt_2\in\T_{L/B}$ and any type of goodness~$\tau$, we have
\begin{equation}
\label{2eta}
\BbbP_{L,\beta}\bigl(\vartheta_{\bt_1}(\GG_\tau)\cap\vartheta_{\bt_2}(\GG_\tau^\cc)\bigr)\le h(\eta),
\end{equation}
provided~$L\ge L_0$, where~$L_0=L_0(\kappa,\eta)$ is as in Theorem~\ref{thm3.1}.
\end{lemma}

\begin{proofsect}{Proof}
Noting that~$\GG_\tau\cap\GG_\tau^\cc=\emptyset$, let us assume that $\bt_1\ne\bt_2$.
Now, for the intersection $\vartheta_{\bt_1}(\GG_\tau)\cap\vartheta_{\bt_2}(\GG_\tau^\cc)$ to occur, either the block at $B\bt_2$ is bad, which has probability at most~$\eta$, or it is good but \emph{not} of the type~$\tau$. We claim that, in the latter case, there must be a ``surface'' consisting of bad blocks which separates the block at~$B\bt_1$ from that in~$B\bt_2$. Indeed, let $\bS\in\vartheta_{\bt_1}(\GG_\tau)$ and consider the connected component,~$\scrC$, of good blocks in~$\T_{L/B}$ containing the block at~$B\bt_1$. We claim that the \emph{type} of goodness is constant throughout~$\scrC$, i.e., it is of type~$\tau$. To see this, suppose that a block in~$\scrC$ has the type of goodness which is distinct from~$\tau$. By the fact that~$\scrC$ is connected, there must exist a pair of \emph{neighboring} blocks with distinct types of goodness. But neighboring blocks \emph{share} the sites on their separating face and (since~$\kappa\ll1$) the spins on this face cannot simultaneously be in the~$2\kappa$-neighborhood of two~$\hatw_\tau$'s---that is, not without the spins busting apart. Hence, on~$\vartheta_{\bt_1}(\GG_\tau)\cap\vartheta_{\bt_2}(\GG_\tau^\cc)$, the block at~$B\bt_2$ is not part of~$\scrC$ and we have it separated from~$B\bt_1$ by a ($*$-connected) ``surface'' of bad~blocks.

To estimate the probability of such a ``surface'' we will use Theorem~\ref{thm3.1}: The probability that a ``surface'' involving altogether~$m$ \emph{given} bad blocks occurs is bounded by~$\eta^m$. The rest of the proof parallels the standard Peierls argument which hinges upon the fact that the number~$N_m$ of $*$-connected ``surfaces'' comprising~$m$ blocks and containing a \emph{given block} grows only exponentially with~$m$, i.e., $N_m\le c^m$ for some~$c\in(1,\infty)$. To count the number of ways how to choose the particular block in the ``surface,'' we have to be a bit cautious about the toroidal geometry: If $m<L/B$, then the ``surface'' encloses either the block at~$B\bt_1$ or that at~$B\bt_2$ on all sides and there are at most~$2m$ ways to choose one particular block. On the other hand, if $m\ge L/B$, then the surface can be topologically non-trivial but, since~$\T_{L/B}$ is a finite graph, the number of choices of one particular block is at most~$(L/B)^3\le m^3$. This shows that \eqref{2eta} holds with
\begin{equation}
h(\eta)=\eta+2\sum_{m\ge6}m^3 (c\eta)^m,
\end{equation}
uniformly in~$L\ge L_0$. Clearly, $h(\eta)\to0$ as $\eta\downarrow0$. 
\end{proofsect}

\smallskip
Now we are ready to prove the existence of long-range order in 120$^\circ$-model, subject to the validity of Theorem~\ref{thm3.1}:

\begin{proofsect}{Proof of Theorem~\ref{thm2.1}}
Let~$\eta>0$ and let~$\beta_0$ and~$L_0$ be as in Theorem~\ref{thm3.1}. Fix a~$\beta\ge\beta_0$ and choose~$B$ and~$\Gamma$ accordingly. For finite~$L\ge L_0$, it follows by \eqref{3.4aa} that, with probability exceeding $1-\eta$, any given block is in a good state. Since the distinct types of goodness are disjoint and related by symmetry, we have
\begin{equation}
\label{3.6}
\BbbP_{L,\beta}\bigl(\vartheta_{\bt}(\GG_\tau)\bigr)\ge\frac16(1-\eta)
\end{equation}
for any~$\bt\in\T_{L/B}$ and any $\tau=1,2,\dots,6$. Next, we may condition the block farthest from the origin (i.e., the one at the ``back'' of the torus) to be of a particular type of goodness, say~$\GG_\tau$. The resulting measure still satisfies the~DLR-condition in any subset of the torus not intersecting the far-away block. Passing to the thermodynamic limit along \emph{some} sequence of~$L$'s, we arrive at an infinite-volume Gibbs state for the interaction~\eqref{Hdiff}. 

Clearly, by \eqref{3.6} and Lemma~\ref{lemma3.2}, we have the uniform bound
\begin{equation}
\BbbP_{L,\beta}\bigl(\vartheta_{\bt_1}(\GG_\tau^\cc)\bigl|\vartheta_{\bt_2}(\GG_\tau)\bigr)
\le6\,\frac{h(\eta)}{1-\eta}.
\end{equation}
Hence, if~$\eta\ll1$, we have constructed six infinite-volume Gibbs states in the~120$^\circ$-model which are distinguished by the statistical properties of \emph{any} individual spin. In particular, the bound \eqref{lbexp} holds with~$\epsilon(\beta)$ directly related to~$\eta$,~$h(\eta)$ and~$\kappa$. Of course, it is not automatically the case that the resulting states are translation-invariant; however, this is easily handled by considering a translation average of the abovementioned and noting that the ``distinctness'' of the states via the single spin observables is preserved by this averaging.
\end{proofsect}

\section{Spin-wave analysis}
\label{sec4}\noindent
This section provides rigorous justification for the heuristic spin-wave calculations from Sect.~\ref{sec2.2}. Beyond the fact that these calculations settle the pertinent questions concerning long-range order at the non-rigorous level, such results, as refined here, serve as the cornerstone for the proof of Theorem~\ref{thm3.1}. The principal results of this section are Theorems~\ref{thm4.1} and~\ref{thm4.10}.

\subsection{Homogeneous ground states}
\label{sec4.1}\noindent
Our goal is to evaluate the free energy of the spin configurations where all spins are more or less aligned with a given vector on the unit circle.
Let us represent all of the spins $\bS=(\bS_{\br})$ by their corresponding angle variables $\btheta=(\theta_{\br})$---vis-a-vis the usual $\bS_{\br}=(\cos\theta_{\br},\sin\theta_{\br})$---and let~$\theta^\star$ denote the particular direction towards which we wish the spins to align. Let~$\chi_{\aD,L}(\btheta)$ be the indicator of the event that $|\theta_{\br}-\theta^\star|<\aD$, with the difference $\theta_{\br}-\theta^\star$ interpreted modulo~$2\pi$, holds for all~$\br\in\T_L$. Here~$\aD$ is closely related to the 
quantity~$\Gamma$ from Sect.~\ref{sec3.2}.

In this representation we define the constrained free energy by the formula
\begin{equation}
\label{FDLB1}
F_{L,\beta}^{(\aD)}(\theta^\star)=
-\frac12\log\frac{\beta J}{2\pi}-\frac1{L^3}\log \int
e^{-\beta\mathscr{H}_L(\btheta)}\chi_{\aD,L}(\btheta)\prod_{\br\in\T_L}\textd\theta_{\br},
\end{equation}
where $\mathscr{H}_L(\btheta)$ denotes the torus Hamiltonian expressed in terms of the angle variables~$\btheta$ and where the first term on the right-hand side has been added for later convenience.
Our goal is to show that, under specific conditions,~$F_{L,\beta}^{(\aD)}(\theta^\star)$ can be well approximated by the function~$F$ defined in~\eqref{Fn1}. (As is easy to check, the integral in \eqref{Fn1} converges for all~$\theta^\star$.)

\smallskip
Recall the abbreviations~$q_1=\sin^2(\theta^\star)$,~$q_2=\sin^2(\theta^\star-\tfrac{2\pi}3)$ and~$q_3=\sin^2(\theta^\star+\tfrac{2\pi}3)$ from Sect.~\ref{sec2.2}.
The precise statement concerning the above approximation is as follows:

\begin{theorem}
\label{thm4.1}
For each $\epsilon>0$ there exists a number $\delta=\delta(\epsilon)>0$
such that if~$\beta J$ and~$\aD$ obey
\begin{equation}
\label{4.3}
(\beta J)\aD^2\ge 1/\delta\quad\text{and}\quad (\beta J)\aD^3\le \delta,
\end{equation}
then 
\begin{equation}
\label{4.5a}
\limsup_{L\to\infty}\,\bigl|F_{L,\beta}^{(\aD)}(\theta^\star)-F(\theta^\star)\bigr|\le 
\epsilon
\end{equation}
for all $\theta^\star\in[0,2\pi)$.
\end{theorem}

As the first step of the proof, we will pass to the harmonic approximation of the Hamiltonian, which is given by
\begin{equation}
\label{ILT}
\mathscr{I}_L(\btheta)=\frac{\beta J}2\sum_{\br\in\T_L}\sum_\alpha q_\alpha(\theta_{\br}-\theta_{\br+\hate_\alpha})^2.
\end{equation}
The next lemma provides an estimate of the error in this approximation.

\begin{lemma}
\label{lemma4.2a}
There exists a constant~$c_1\in(0,\infty)$ such that for all $\beta\in(0,\infty)$, all $\aD\in(0,1)$, all $L\ge1$ and all $\theta^\star\in[0,2\pi)$ the following holds:
If $\chi_{\aD,L}(\btheta)=1$, then
\begin{equation}
\bigl|\beta\mathscr{H}_L(\btheta)-\mathscr{I}_L(\btheta)\bigr|\le c_1(\beta J)\aD^3L^3.
\end{equation}
\end{lemma}

\begin{proofsect}{Proof}
Let us first consider the nearest-neighbor bond $(\br,\br+\hate_1)$ and note that $S_{\br}^{(1)}=\cos\theta_{\br}$. Since $|\theta_{\br}-\theta^\star|\le\aD$, Taylor's Theorem gives us the bound
\begin{equation}
\bigl|S_{\br}^{(1)}-S_{\br+\hate_1}^{(1)}+\sin(\theta^\star)(\theta_{\br}-\theta_{\br+\hate_1})\bigr|\le \aD^2.
\end{equation}
But $|\theta_{\br}-\theta_{\br+\hate_1}|\le2\aD$ and thus $(S_{\br}^{(1)}-S_{\br+\hate_1}^{(1)})^2$ and~$q_1(\theta_{\br}-\theta_{\br+\hate_1})^2$ differ by less than a numerical constant times~$\aD^3$.
The situation in the other directions is similar, one just has to note that $S_{\br}^{(2)}=\cos(\theta-\tfrac{2\pi}3)$ and $S_{\br}^{(3)}=\cos(\theta+\tfrac{2\pi}3)$.
Adding up the contribution of all three components, multiplying by~$\beta J$ and summing over $\br\in\T_L$, the result directly follows.
\end{proofsect}

Having converted the Boltzmann weight $e^{-\beta\mathscr{H}_L(\bS)}$ into the Gaussian weight $e^{-\mathscr{I}_L(\btheta)}$ in \eqref{FDLB1}, our next task is to estimate the effect of the indicator~$\chi_{\aD,L}$. Let
\begin{equation}
\label{QLD1}
Q_{L,\aD}^{(\theta^\star\!,\beta)}=\Bigl(\frac{\beta J}{2\pi}\Bigr)^{L^3/2}\int e^{-\mathscr{I}_L(\btheta)}\chi_{\aD,L}(\btheta)\prod_{\br\in\T_L}\textd\theta_{\br}.
\end{equation}
Then we have:

\begin{lemma}
\label{lemma4.3a}
For all $\beta\in(0,\infty)$, all $\aD\in(0,1)$ and all $\theta^\star\in[0,2\pi)$,
\begin{equation}
\label{Qlim1}
\limsup_{L\to\infty}\,
\frac{\log Q_{L,\aD}^{(\theta^\star\!,\beta)}}{L^3}\le -F(\theta^\star).
\end{equation}
\end{lemma}

\begin{proofsect}{Proof}
We will use the exponential Chebyshev inequality. Let $\lambda>0$. Then the indicator~$\chi_{\aD,L}$ is bounded via
\begin{equation}
\label{chbd120}
\chi_{\aD,L}(\btheta)\le
e^{\frac12\lambda(\beta J)\aD^2L^3}\exp\Bigl\{-\frac12\lambda(\beta J)
\sum_{\br\in\T_L}(\theta_{\br}-\theta^\star)^2\Bigr\}.
\end{equation}
Plugging the right-hand side into \eqref{QLD1} instead of $\chi_{\aD,L}$, we get a Gaussian integral with $L^3$-dimensional covariance matrix $C=(\beta J)^{-1}(\lambda\1+\widehat D)^{-1}$, where~$\1$ is the unit matrix and~$\widehat D$ is a generalized Laplacian implicitly defined by~\eqref{ILT}.
Integrating out the variables~$\btheta$ and invoking Fourier transform to diagonalize~$C$, we get
\begin{equation}
\label{4.11a}
\frac{\log Q_{L,\aD}^{(\theta^\star\!,\beta)}}{L^3}\le\frac12\lambda(\beta J)\aD^2
-\frac12\frac1{L^3}\sum_{\bk\in\T_L^\star}
\log\Bigl\{\lambda+\sum_\alpha q_\alpha|1-e^{\texti k_\alpha}|^2\Bigr\},
\end{equation}
where $\T_L^\star$ denotes the reciprocal volume (or the Brillouin zone).
Passing to the limit $L\to\infty$, we find out that the left-hand side of \eqref{Qlim1} is bounded by $\frac12\lambda(\beta J)\aD^2-F(\theta^\star,\lambda)$, where
\begin{equation}
\label{FTL}
F(\theta^\star,\lambda)=\frac12\int_{[-\pi,\pi]^3}
\frac{\textd\bk}{(2\pi)^3}
\log\Bigl\{\lambda+\sum_\alpha q_\alpha|1-e^{\texti k_\alpha}|^2\Bigr\}.
\end{equation}
But the integrand is a monotone function of~$\lambda$, and so the Monotone Convergence Theorem guarantees that $F(\theta^\star,\lambda)\downarrow F(\theta^\star)$ as $\lambda\downarrow0$. Thence the result follows by taking~$\lambda$ to zero.
\end{proofsect}

Let $F(\theta^\star,\lambda)$ be the quantity defined in \eqref{FTL}. The lower bound is then as follows:

\begin{lemma}
\label{lemma4.5a}
For all $\beta\in(0,\infty)$, all $\aD\in(0,1)$, all $\theta^\star\in[0,2\pi)$, and all $\lambda>0$ satisfying $(\beta J)\aD^2\lambda>1$,
\begin{equation}
\liminf_{L\to\infty}\,
\frac{\log Q_{L,\aD}^{(\theta^\star\!,\beta)}}{L^3}\ge -F(\theta^\star,\lambda)
+\log\Bigl(1-\frac1{\beta J\aD^2}\frac1\lambda\Bigr).
\end{equation}
\end{lemma}

\begin{proofsect}{Proof}
Let $\lambda>0$ and consider the Gaussian measure~$\BbbP_\lambda$ given by
\begin{equation}
\BbbP_\lambda(\textd\btheta)=
\frac1{Q_{L,\theta^\star}}\Bigl(\frac{\beta J}{2\pi}\Bigr)^{L^3/2}
\exp\biggl\{-\mathscr{I}_L(\btheta)-\frac12\lambda(\beta J)
\sum_{\br\in\T_L}(\theta_{\br}-\theta^\star)^2\biggr\}
\prod_{\br\in\T_L}\textd\theta_{\br},
\end{equation}
where~$Q_{L,\theta^\star}$ is the corresponding normalization factor, which modulo the ``log'' and the factor~$L^3$ equals the final term on the right-hand side of \eqref{4.11a}.
Let us use $\E_\lambda$ to denote the corresponding expectation. Then
\begin{equation}
\label{4.25a}
Q_{L,\aD}^{(\theta^\star\!,\beta)}=Q_{L,\theta^\star}\,\E_\lambda(\chi_{\aD,L}).
\end{equation}
Now $\chi_{\aD,L}$ is simply the product of indicators of the type $\1_{\{|\theta_{\br}-\theta^\star|<\aD\}}$. We claim that
\begin{equation}
\label{4.26a}
\E_\lambda(\chi_{\aD,L})=\E_\lambda\Bigl(\prod_{\br\in\T_L}\1_{\{|\theta_{\br}-\theta^\star|<\aD\}}\Bigr)
\ge\prod_{\br\in\T_L}\BbbP_\lambda\bigl(|\theta_{\br}-\theta^\star|<\aD\bigr).
\end{equation}
This follows from the fact that the moduli of these sorts of Gaussian fields are FKG-positively correlated, see e.g.~\cite{BCG}. It can also be established on the basis of the ``esoteric'' version of reflection positivity (using reflections between sites), which is described at the beginning of Sect.~\ref{sec3.1}.
The estimate thus boils down to a lower bound on the probability of $|\theta_{\br}-\theta^\star|<\aD$.

Now, let us note that the Fourier components~$\widehat\theta_{\bk}$ of the fields~$\theta_{\br}-\theta^\star$ have $\E_\lambda(\widehat\theta_{\bk})=0$ and, for~$\bk'\ne\pm\bk$, the random variables $\widehat\theta_{\bk}$ and~$\widehat\theta_{\bk'}$ are independent with 
\begin{equation}
\E_\lambda\bigl(|\widehat\theta_{\bk}|^2\bigr)=\frac1{\beta J}
\Bigl(\lambda+\sum_\alpha q_\alpha|1-e^{\texti k_\alpha}|^2\Bigr)^{-1}\le\frac1{\beta J\lambda}.
\end{equation}
Thus, invoking the Chebyshev inequality the complementary probability is bounded by
\begin{equation}
\label{4.28a}
\BbbP_\lambda\bigl(|\theta_{\br}-\theta^\star|\ge\aD\bigr)
\le\frac{\E_\lambda(|\theta_{\br}-\theta^\star|^2)}{\aD^2}
=\frac1{L^3}\sum_{\bk\in\T_L^\star}\frac{\E_\lambda(|\widehat\theta_{\bk}|^2)}{\aD^2}\le
\frac1{\beta J\aD^2}\frac1\lambda. 
\end{equation}
Combining \twoeqref{4.25a}{4.26a} with \eqref{4.28a}, invoking the explicit expression for~$Q_{L,\theta^\star}$ and passing to the limit~$L\to\infty$, the desired bound is proved.
\end{proofsect}

Now we are ready to prove the error bound in \eqref{4.5a}:

\begin{proofsect}{Proof of Theorem~\ref{thm4.1}}
By the Monotone Convergence Theorem we have that $F(\theta^\star,\lambda)\downarrow F(\theta^\star)$ as $\lambda\downarrow0$. Moreover, the continuity of $\theta^\star\mapsto F(\theta^\star)$ and the fact that the unit circle in~$\R^2$ is compact imply that this convergence is actually uniform in~$\theta^\star$. Hence, for each $\epsilon>0$, there exists a number $\lambda>0$ such~that
\begin{equation}
\label{2Fa}
\bigl|F(\theta^\star,\lambda)-F(\theta^\star)\bigr|\le\frac\epsilon3
\end{equation}
for $\theta^\star\in[0,2\pi)$. 
Let~$c_1$ be the constant from Lemma~\ref{lemma4.2a} and choose~$\delta$ such that $c_1\delta\le\epsilon/3$. Suppose also that~$\delta<\lambda$ and
\begin{equation}
\log\Bigl(1-\frac\delta\lambda\Bigr)\ge-\frac\epsilon3.
\end{equation}
Fix an angle~$\theta^\star\in[0,2\pi)$.
Lemma~\ref{lemma4.2a} along with our choice of~$\delta$ imply
\begin{equation}
\limsup_{L\to\infty}\,\Bigl|\frac{\log Q_{L,\aD}^{(\theta^\star,\beta)}}{L^3}
+F_{L,\beta}^{(\aD)}(\theta^\star)\Bigr|\le\frac\epsilon3.
\end{equation}
On the other hand, Lemmas~\ref{lemma4.3a}--\ref{lemma4.5a}, the choice of~$\lambda$ in \eqref{2Fa} and our choice of~$\delta$ ensure that
\begin{equation}
\limsup_{L\to\infty}\,\Bigl|\frac{\log Q_{L,\aD}^{(\theta^\star,\beta)}}{L^3}
+F(\theta^\star)\Bigr|\le\frac23\epsilon.
\end{equation}
Combining these two estimates, the bound \eqref{4.5a} is proved.
\end{proofsect}

\subsection{Stratified ground states}
\label{sec4.3}\noindent
As mentioned previously, constant configurations are only the overture for the set of all possible ground states. As a consequence, the knowledge of the spin-wave free energy about homogeneous background configurations is not sufficient for the proofs of our main results. Fortunately, as we shall see in Sect.~\ref{sec6}, the chessboard estimates allow us to reduce the (potentially quite large) number of remaining cases to configurations which are translation-invariant in two directions and alternating in the third direction. 

To avoid parity problems, throughout this section we will assume that~$L$ is an even integer.
Fix an index~$\alpha\in\{1,2,3\}$, pick a direction~$\theta^\star\in[0,2\pi)$ and let~$\tilde\theta^\star$ denote the reflection of~$\theta^\star$ through the~$\alpha$-th of the vector~$\hata$,~$\hatb$ or~$\hatc$. Consider again the angle variables~$\theta_{\br}$ and let~$\widetilde\chi_{\aD,L}$ be the indicator that~$|\theta_{\br}-\theta^\star|<\aD$ for~$\br\in\T_L$ with an even~$\alpha$-th component while for~$\br$ with an odd~$\alpha$-component we require that~$|\theta_{\br}-\tilde\theta^\star|<\aD$. Let
\begin{equation}
\label{FDLB2}
\widetilde F_{L,\beta}^{(\aD,\alpha)}(\theta^\star)=
-\frac12\log\frac{\beta J}{2\pi}-\frac1{L^3}\log \int
e^{-\beta\mathscr{H}_L(\btheta)}
\widetilde\chi_{\aD,L}(\btheta)\prod_{\br\in\T_L}\textd\theta_{\br}.
\end{equation}
The quantity $\widetilde F_{L,\beta}^{(\aD,\alpha)}(\theta^\star)$ represents the spin-wave free energy for (period-two) stratified states perpendicular to direction~$\alpha$ and spins alternating between directions~$\theta^\star$ and~$\tilde\theta^\star$.

As before, our goal is to approximate $\widetilde F_{L,\beta}^{(\aD,\alpha)}(\theta^\star)$ by an appropriate momentum-space integral. For~$\alpha\in\{1,2,3\}$, let us abbreviate
\begin{equation}
E_\alpha=E_\alpha(\bk)=|1-e^{\texti k_\alpha}|^2
\quad\text{and}\quad
E_\alpha^\star=E_\alpha^\star(\bk)=|1+e^{\texti k_\alpha}|^2,
\end{equation}
where~$k_\alpha$ is the~$\alpha$-th component of the vector~$\bk\in[-\pi,\pi]^3$, and recall, once again, the meaning of the quantities~$q_\alpha$ (cf Sect.~\ref{sec2.2}). We will define three $2\times2$-matrices $\Pi_\alpha(\bk)$, $\alpha=1,2,3$. First let $\alpha=1$ and abbreviate~$q_+=\frac12(q_2+q_3)$ and~$q_-=\frac12(q_2-q_3)$. Then 
\begin{equation}
\Pi_1(\bk)=
\left(
\begin{matrix}
q_1 E_1+q_+(E_2+E_3) &  q_-(E_2-E_3)\\
q_-(E_2-E_3) & q_1 E_1^\star+q_+(E_2+E_3)
\end{matrix}\right).
\end{equation}
The quantities~$\Pi_2$ and~$\Pi_3$ are defined by cyclically permuting the roles of $E_1$, $E_2$ and~$E_3$ and similarly for the~$q_\alpha$'s. (In the physically relevant quantities,~$q_-$ will appear only in terms of its square, so the order used for the definition of this quantity is for all intents and purposes arbitrary.)
Then we define a function~$\widetilde F_\alpha$ assigning to each $\theta^\star\in[0,2\pi)$ and each~$\alpha\in\{1,2,3\}$ the value
\begin{equation}
\label{4.56}
\widetilde F_\alpha(\theta^\star)=\frac14\int_{[-\pi,\pi]^3}
\frac{\textd\bk}{(2\pi)^3}\log\det \Pi_\alpha(\bk).
\end{equation}
The fact that now we have a quarter in front of the integral comes from the fact that the determinant actually represents the combined contribution of two~$\bk$-modes.

\smallskip
The main result of this section concerning the 120$^\circ$-model is now as follows:

\begin{theorem}
\label{thm4.10}
For each $\epsilon>0$ there exists a number $\delta=\delta(\epsilon)>0$
such that if~$\beta J$ and~$\aD$ obey
\begin{equation}
\label{4.3s}
(\beta J)\aD^3\le \delta,
\end{equation}
then 
\begin{equation}
\label{4.5as}
\liminf_{L\to\infty}\,\widetilde F_{L,\beta}^{(\aD,\alpha)}(\theta^\star)\ge \widetilde F_\alpha(\theta^\star)-\epsilon
\end{equation}
for all $\theta^\star\in[0,2\pi)$ and all $\alpha\in\{1,2,3\}$.
\end{theorem}

As we have seen in the previous sections, the first step is to pass to the quadratic approximation of the torus Hamiltonian. Let $\btheta=(\theta_{\br})$ be a configuration of angle variables. Then we define
\begin{equation}
\label{ILTs}
\widetilde{\mathscr{I}}_{L,\alpha}(\btheta)=\frac{\beta J}2\sum_{\br\in\T_L}\sum_{\gamma=1,2,3} q_{\gamma,\br}^{(\alpha)}(\theta_{\br}-\theta_{\br+\hate_\gamma})^2.
\end{equation}
Here~$q_{\gamma,\br}^{(\alpha)}$ is the usual~$q_\gamma$ if the $\alpha$-th component of~$\br$ is even while for the complementary~$\br$ we have to interchange the roles of the two $q_{\gamma'}$ with~$\gamma'\ne\alpha$. (In particular,~$q_{\alpha,\br}^{(\alpha)}=q_\alpha$ for all~$\br$.)

\smallskip
Our first item of concern is the error caused by this approximation:

\begin{lemma}
\label{lemma4.2as}
There exists a constant~$c_2\in(0,\infty)$ such that for all $\beta\in(0,\infty)$, all $\aD\in(0,1)$, all $L\ge1$ and all $\theta^\star\in[0,2\pi)$ the following holds:
If $\widetilde\chi_{\aD,L}(\btheta)=1$ and if~$\tilde\btheta=(\tilde\theta_{\br})$ is the configuration obtained by reflecting~$\theta_{\br}$ through the~$\alpha$-th of the vectors~$\hata$,~$\hatb$ or~$\hatc$ for~$\br$ with an odd~$\alpha$-component, then
\begin{equation}
\bigl|\beta\mathscr{H}_L(\btheta)-\widetilde{\mathscr{I}}_{L,\alpha}(\tilde\btheta)\bigr|\le c_2(\beta J)\aD^3L^3.
\end{equation}
\end{lemma}

\begin{proofsect}{Proof}
Once we have accounted for the inhomogeneity of the setup, the proof is essentially identical to that of Lemma~\ref{lemma4.2a}. Without much loss of generality, let us focus on the case~$\alpha=1$. Let~$\btheta$ be such that $\widetilde\chi_{\aD,L}(\btheta)=1$ and let~$\tilde\btheta$ be as described. 

We will concentrate on the interaction of two spins in the~$y$-coordinate direction.
If the~$\alpha$-component of~$\br$ is even, then the expansion around~$\theta_{\br}=\tilde\theta_{\br}\approx\theta^\star$ gives that $\cos(\theta_{\br}-\frac{2\pi}3)$ is well approximated by~$\cos(\theta^\star-\frac{2\pi}3)-\sin(\theta^\star-\frac{2\pi}3)(\tilde\theta_{\br}-\theta^\star)$. Accounting better for the errors we thus~get
\begin{equation}
\bigl|S_{\br}^{(2)}-S_{\br+\hate_2}^{(2)}+\sin\bigl(\theta^\star-\tfrac{2\pi}3\bigr)(\tilde\theta_{\br}-\tilde\theta_{\br+\hate_2})\bigr|\le \aD^2.
\end{equation}
On the other hand, for~$\br$ with an odd $\alpha$-th component we have~$-\theta_{\br}=\tilde\theta_{\br}\approx\theta^\star$ which means that $S_{\br}^{(2)}=\cos(-\tilde\theta_{\br}-\frac{2\pi}3)=\cos(\tilde\theta_{\br}+\frac{2\pi}3)$ and thus
\begin{equation}
\bigl|S_{\br}^{(2)}-S_{\br+\hate_2}^{(2)}-\sin\bigl(\theta^\star+\tfrac{2\pi}3\bigr)
(\tilde\theta_{\br}-\tilde\theta_{\br+\hate_2})\bigr|\le \aD^2.
\end{equation}
After plugging into~$\mathscr{H}_L(\btheta)$, the~$\theta_{\br}$ in the even~$\br$ planes are coupled by~$q_2$ while in the odd~$\br$ planes they are coupled by~$q_3$, in accord with~\eqref{ILTs}.

A completely analogous argument handles the case of two sites in the~$z$-coordinate direction. In the~$x$-direction the reflection has no effect because the minus sign from $\sin(-\theta^\star)$ disappears after we take the square. The ensuing errors are estimated exactly as in the proof of Lemma~\ref{lemma4.2a}.
\end{proofsect}

Since~$\widetilde\chi_{\aD,L}(\btheta)=\chi_{\aD,L}(\tilde\btheta)$, where~$\btheta$ and~$\tilde\btheta$ are related as in Lemma~\ref{lemma4.2as}, a simple change of variables shows that the proper analogue of the quantity from \eqref{QLD1} for the present setup is 
\begin{equation}
\label{QLD1b}
\widetilde Q_{L,\aD,\beta}^{(\theta^\star\!,\alpha)}=\Bigl(\frac{\beta J}{2\pi}\Bigr)^{L^3/2}\int e^{-\widetilde{\mathscr{I}}_{L,\alpha}(\btheta)}\chi_{\aD,L}(\btheta)\prod_{\br\in\T_L}\textd\theta_{\br}.
\end{equation}
Note that here the inhomogeneity of the domain of integration in \eqref{FDLB2} has now been moved to the Gaussian weight. Next we apply:

\begin{lemma}
\label{lemma4.3as}
For all $\beta\in(0,\infty)$, all $\aD\in(0,1)$, all $\theta^\star\in[0,2\pi)$ and all $\alpha\in\{1,2,3\}$,
\begin{equation}
\label{Qlim1s}
\limsup_{L\to\infty}\,
\frac{\log \widetilde Q_{L,\aD,\beta}^{(\theta^\star\!,\alpha)}}{L^3}\le -\widetilde F_\alpha(\theta^\star).
\end{equation}
\end{lemma}

\begin{proofsect}{Proof}
Fix~$\alpha$ and let~$\lambda>0$. The proof again commences by invoking the exponential Chebyshev inequality in the form of \eqref{chbd120}. The resulting $L^3$-dimensional Gaussian integral has covariance matrix $C_\alpha=(\beta J)(\lambda\1+\widehat D^{(\alpha)})$, where~$\frac12\beta J\widehat D^{(\alpha)}$ is the matrix corresponding to the quadratic form~\eqref{ILTs} in the variables~$\theta_{\br}$. The difference compared to Lemma~\ref{lemma4.3a} is that now~$\widehat D^{(\alpha)}$ is no longer translation-invariant in the~$\alpha$-th direction, but only periodic with period two. As a result, the~$\bk$ and $\bk+\pi\hate_\alpha$ modes will mix together and the Fourier transform of~$C_\alpha$ will result in $2\times2$-block-diagonal matrix. The blocks are parametrized by pairs of momenta $(\bk,\bk+\pi\hate_\alpha)$. (Note that, since~$L$ is even, $\bk+\pi\hate_\alpha\in\T_L^\star$ whenever~$\bk\in\T_L^\star$.)

A calculation---which is best performed by taking the Fourier transform of $\widetilde{\mathscr{I}}_{L,\alpha}$---reveals that the block corresponding to the pair $(\bk,\bk+\pi\hate_\alpha)$ is exactly~$\Pi_\alpha(\bk)$. Hence we get
\begin{equation}
\label{4.11as}
\frac{\log \widetilde Q_{L,\aD,\beta}^{(\theta^\star\!,\alpha)}}{L^3}
\le\frac12\lambda(\beta J)\aD^2
-\frac14\frac1{L^3}\sum_{\bk\in\T_L^\star}
\log\det\bigl(\lambda\1+\Pi_\alpha(\bk)\bigr),
\end{equation}
where~$\1$ is the $2\times2$-unit matrix and where the usual factor~$1/2$ in front of the sum is replaced by a~$1/4$ to account for the fact that $\bk$ and $\bk+\pi\hate_\alpha$ are treated as independent entities in the sum. (We are using the $\bk\leftrightarrow\bk+\pi\hate_\alpha$ symmetry of the determinant.) Passing to limits $L\to\infty$ and $\lambda\downarrow0$, the bound \eqref{Qlim1s} is proved.
\end{proofsect}

\begin{proofsect}{Proof of Theorem~\ref{thm4.10}}
By Lemma~\ref{lemma4.2as} and the definition of $\widetilde F_{L,\beta}^{(\aD,\alpha)}(\theta^\star)$ we know that
\begin{equation}
\widetilde F_{L,\beta}^{(\aD,\alpha)}(\theta^\star)\ge 
-\frac{\log \widetilde Q_{L,\aD,\beta}^{(\theta^\star\!,\alpha)}}{L^3}-c_2(\beta J)\aD^3.
\end{equation}
Hence, if~$\delta$ is such that $c_2\delta\le\epsilon$, \eqref{4.5as} follows by taking $L\to\infty$ and invoking Lemma~\ref{lemma4.3as}.
\end{proofsect}

\section{Spin-wave free energy minima}
\label{sec5}\noindent
The purpose of this section is to show that the spin-wave free energy $F(\theta^\star)$, which emerges from the analysis in Sect.~\ref{sec4.1}, is minimized in the ``directions'' as stated in Theorems~\ref{thm2.1}. Similarly, we will also show that the free energy $\widetilde F_\alpha(\theta^\star)$ corresponding to the inhomogeneous ground states is always strictly larger than its homogeneous counterpart $F(\theta^\star)$, unless~$\theta^\star$ is ``aligned'' (or ``antialigned'') with the~$\alpha$-th of the vectors~$\hata$,~$\hatb$ or~$\hatc$. These findings constitute the essential ingredients for the proof of Theorem~\ref{thm3.1} in Sect.~\ref{sec6.2}. The principal estimates are based on Jensen's inequality combined with a non-trivial bit of ``function analysis.''

\subsection{Homogeneous ground states}
\label{sec5.1}\noindent
Our task is to identify the minima of the function $\theta^\star\mapsto F(\theta^\star)$ defined in \eqref{Fn1}. However, noting that the product structure of the measure $\textd\bk/(2\pi)^3$ makes the random variables $|1-e^{\texti k_\alpha}|^2$ independent, we might as well analyze an entire class of functions of this type.

\smallskip
Let $X$ be a random variable taking values in $[-1,1]$ and, for any triple of numbers $(a,b,c)$, define the function
\begin{equation}
F(a,b,c)=\E\Bigl(\log\bigr(a^2(1-X_1)+b^2(1-X_2)+c^2(1-X_3)\bigr)\Bigr),
\end{equation}
where $X_1,X_2$ and $X_3$ are independent copies of~$X$.
Suppose in addition that the distribution~$\mu$ of~$X$ has the following properties
\begin{enumerate}
\item[(1)]
$\mu$ has a density with respect to the Lebesgue measure, $\mu(\textd x)=f(x)\textd x$.
\item[(2)]
$f(x)=f(-x)$ for all $x\in[-1,1]$.
\item[(3)]
$f(x)$ is strictly increasing on $[0,1]$.
\end{enumerate}
Then we have the following general result:

\begin{theorem}
\label{thm5.1}
Let $a,b,c\mapsto F(a,b,c)$ be as above with~$X$ satisfying the properties (1-3).
Then for any nonzero $\varkappa\in\R$ and any $a,b,c$ satisfying
\begin{equation}
\label{abc-constraints}
a+b+c=0
\quad\text{and}\quad
a^2+b^2+c^2=2\varkappa^2,
\end{equation}
we have
\begin{equation}
\label{Fbd}
F(a,b,c)\ge F(0,\varkappa,-\varkappa).
\end{equation}
Moreover, the inequality is strict whenever $a,b,c\ne0$ and $F(0,\varkappa,-\varkappa)>-\infty$.
\end{theorem}

The particular case of the 120$^\circ$-model can now be easily extracted:

\begin{corollary}
\label{cor5.2}
The function $\theta^\star\mapsto F(\theta^\star)$ achieves its minimum only at the points
\begin{equation}
\label{thval}
\theta^\star=\frac\pi3\tau,\qquad \tau=1,2,\dots,6.
\end{equation}
\end{corollary}

\begin{proofsect}{Proof}
We just have to identify the quantities $a,b,c$ and the random variable~$X$ in the case of the 120$^\circ$-model. First, since $|1-e^{\texti k_\alpha}|^2=2(1-\cos k_\alpha)$, we let $X$ be the random variable distributed as $\cos k$ in measure~$\textd k/(2\pi)$ on $[-\pi,\pi]$. A simple calculation shows that~$X$ has a density $f(x)=(1-x^2)^{-1/2}/(2\pi)$ with respect to the Lebesgue measure on $[-1,1]$, which manifestly satisfies the requirements~(2--3) above. 

Now, setting $a=\sqrt2\sin(\theta^\star)$, $b=\sqrt2\sin(\theta^\star-\tfrac{2\pi}3)$ and $c=\sqrt2\sin(\theta^\star+\tfrac{2\pi}3)$, we have that $F(a,b,c)=2F(\theta^\star)$. Moreover, a trivial calculation shows that $a+b+c=0$, while $a^2+b^2+c^2=3$ and \eqref{abc-constraints} thus holds with $\varkappa^2=3/2$. As a consequence, $\theta^\star\mapsto F(\theta^\star)$ is minimized only by~$\theta^\star$ such that one of the numbers $a,b,c$ vanishes. This is easily checked to give just the values in \eqref{thval}
\end{proofsect}

The rest of this section is devoted to the proof of Theorem~\ref{thm5.1}.
The proof is based on two observations: First, a lemma due to Pearce~\cite{Pearce}:

\begin{lemma}
\label{lemmaPearce}
Let $X$ be a random variable on $[-1,1]$ satisfying properties (1-3) above. For each $\lambda\in\R$, let $\langle-\rangle_\lambda$ denote the expectation with respect to the probability measure $\omega_\lambda(\textd x)=N_\lambda f(x)e^{\lambda x}\textd x$, where~$f$ is the probability density of~$X$ and~$N_\lambda$ is an appropriate normalization constant. Then the function $\lambda\mapsto\langle X\rangle_\lambda$ is strictly concave on $[0,\infty)$.
\end{lemma}

\begin{proofsect}{Proof}
See~\cite{Pearce}.
The conditions~(1-3) represent one of the general situations in which one can prove the~GHS inequality in lattice spin systems, see~\cite[Theorem~II.13.5(iii)]{Simon}.
\end{proofsect}

The second observation is that the constraints \eqref{abc-constraints} reproduce themselves, rather unexpectedly, at the level of quartic polynomials in~$a$,~$b$ and~$c$:

\begin{lemma}
\label{lemma-abc}
Let $a,b,c$ be numbers satisfying \eqref{abc-constraints}. Then
\begin{equation}
a^4+b^4+c^4=2\varkappa^4.
\end{equation}
\end{lemma}

\begin{proofsect}{Proof}
Since $a=-(b+c)$, eliminating $a$ from the second constraint in \eqref{abc-constraints} results in
\begin{equation}
b^2+c^2+bc=\varkappa^2.
\end{equation}
Squaring we get
\begin{equation}
b^4+c^4+b^2c^2+2b^2c^2+2bc(b^2+c^2)=\varkappa^4,
\end{equation}
which can be recast into the form
\begin{equation}
2b^4+2c^4+6b^2c^2+4b^3c+4bc^3=2\varkappa^4.
\end{equation}
Splitting off the term $b^4+c^4$, the rest of the left hand side is clearly $(b+c)^4=a^4$.
\end{proofsect}

With these lemmas in the hand, the proof of Theorem~\ref{thm5.1} is relatively straightforward:

\begin{proofsect}{Proof of Theorem~\ref{thm5.1}}
Since we can scale~$a$,~$b$ and~$c$ by any constant at the cost of changing $F(a,b,c)$ only by an additive $\varkappa$-dependent factor, let us suppose without loss of generality that $\varkappa=1/\sqrt2$. Moreover, if one of $a,b,c$ is zero, say $a=0$, then $b=-c=\pm\varkappa$ and \eqref{Fbd} is trivial. Hence, we only need to focus on the situations when $a,b,c\ne0$.

The first step of the proof is to convert the expectation of the logarithm into the expectation of an exponential. This can be done for instance by invoking the identity
\begin{equation}
\label{5.9}
-\log(1-x)=\int_0^1\textd t\int_0^\infty\textd s\,\frac{e^{-s}}t(e^{stx}-1),
\qquad x<1,
\end{equation}
where the double integral on the right-hand side is well defined because everything is positive.
Let us now plug in $a^2X_1+b^2X_2+c^2X_3$ for~$x$ and take expectation with respect to~$X_1$, $X_2$ and~$X_3$. Applying Fubini's theorem (and the fact that, almost surely, $a^2X_1+b^2X_2+c^2X_3<1$), the result~is
\begin{equation}
F(a,b,c)=\int_0^1\textd t\int_0^\infty\textd s\frac{e^{-s}}t\bigl(1-G(st;a,b,c)\bigr),
\end{equation}
where
\begin{equation}
G(\lambda;a,b,c)=\E\bigl(e^{\lambda(a^2X_1+b^2X_2+c^2X_3)}\bigr).
\end{equation}
We will show that, whenever $a,b,c\ne0$, we have $G(\lambda;a,b,c)<G(\lambda;0,1/\sqrt2,-1/\sqrt2)$ for all $\lambda>0$, from which \eqref{Fbd} and the ensuing conclusion directly follow.

Consider the function $\lambda\mapsto R(\lambda)$ defined by
\begin{equation}
R(\lambda)=\log\frac{G(\lambda;a,b,c)}{G(\lambda;0,1/\sqrt2,-1/\sqrt2)}.
\end{equation}
Our goal is to prove that $R(\lambda)<0$ whenever $\lambda>0$.
First we note that $R(0)=0$ so it suffices to show that $R'(\lambda)<0$ for all $\lambda>0$. Invoking the independence of $X_1$, $X_2$ and $X_3$, we have
\begin{equation}
R'(\lambda)=a^2\langle X\rangle_{\lambda a^2}+b^2\langle X\rangle_{\lambda b^2}+c^2\langle X\rangle_{\lambda c^2}-\langle X\rangle_{\lambda/2},
\end{equation}
where we adhere to the notation from Lemma~\ref{lemmaPearce}.
Now, by Lemma~\ref{lemmaPearce} and our assumptions on the random variable~$X$, the function $\lambda\mapsto\langle X\rangle_\lambda$ is strictly concave. Since $a^2+b^2+c^2=1$ and, as guaranteed by Lemma~\ref{lemma-abc}, also $a^4+b^4+c^4=1/2$, the bound $R'(\lambda)<0$ for $\lambda>0$ is a direct consequence of Jensen's inequality.
\end{proofsect}

\begin{remark}
The previous proof has one (arguably) unnatural feature; namely, the conversion ``from logs to exponentials'' via the identity \eqref{5.9}. It would of some interest (at least for the authors) to see whether a more direct argument can be constructed.
\end{remark}

\subsection{Stratified ground states}
Having identified the absolute minima of the spin-wave free energies for homogeneous background configurations we turn our attention to the free energies corresponding to inhomogeneous ground states. Specifically, we will show that (truly) stratified states have always worse free energy than the corresponding homogeneous ones.

\smallskip
Let $F(\theta^\star)$ denote the spin-wave free-energy from Sect.~\ref{sec4.1} and let $\widetilde F_\alpha(\theta^\star)$ be the corresponding quantity for the stratified states as defined in Sect.~\ref{sec4.3}. Then we have:

\begin{theorem}
\label{thm5.3}
For each~$\kappa>0$ there exists a constant $c_3=c_3(\kappa)>0$ such that if~$\alpha\in\{1,2,3\}$ and if the angle between $\theta^\star\in[0,2\pi)$ and the~$\alpha$-th of the vectors~$\hata$,~$\hatb$ and~$\hatc$ is in~$(\kappa,\pi-\kappa)$, then
\begin{equation}
\widetilde F_\alpha(\theta^\star)\ge F(\theta^\star)+c_3.
\end{equation}
\end{theorem}

\begin{proofsect}{Proof}
Recall the notations $E_\alpha=|1-e^{\texti k_\alpha}|^2$, $E_\alpha^\star=|1+e^{\texti k_\alpha}|^2$ and $q_1=\sin^2(\theta^\star)$, $q_2=\sin^2(\theta-\tfrac{2\pi}3)$ and $q_3=\sin^2(\theta+\tfrac{2\pi}3)$ and the definition of~$\Pi_\alpha(\bk)$ in  \eqref{4.56}.
We will write $\det\Pi_\alpha(\bk)$ as a convex combination of two terms each of which produces the same free energy. Without loss of generality, let us assume that $\alpha=1$. We claim that for all $\bk\in[-\pi,\pi]^3$, the quantity~$\Pi_1(\bk)$ admits the decomposition
\begin{multline}
\label{Pieq}
\qquad
\det\Pi_1(\bk)=\frac12\bigl(q_1E_1+q_2E_2+q_3E_3\bigr)\bigl(q_1E_1^\star+q_3E_2+q_2E_3\bigr)
\\+\frac12\bigl(q_1E_1+q_3E_2+q_2E_3\bigr)\bigl(q_1E_1^\star+q_2E_2+q_3E_3\bigr).
\qquad
\end{multline}
To prove this let us abbreviate~$q_\pm=\frac12(q_2\pm q_3)$ and $E_\pm=E_2\pm E_3$. Focusing on the first term on the right-hand side, we write
\begin{equation}
\begin{aligned}
q_1E_1+q_2E_2+q_3E_3&=q_1E_1+q_+E_++q_-E_-,
\\
q_1E_1^\star+q_3E_2+q_2E_3&=q_1E_1^\star+q_+E_+-q_-E_-,
\end{aligned}
\end{equation}
and similarly for the other two terms. Multiplying these two lines tells us that the first term on the right-hand side of \eqref{Pieq} equals a half of
\begin{equation}
\label{5.26}
(q_1E_1+q_+E_+)(q_1E_1^\star+q_+E_+)-q_-^2E_-^2 +q_1q_-(E_1^\star-E_1)E_-.
\end{equation}
The sole effect of the second product on the right-hand side of \eqref{Pieq} is to cancel the very last term of \eqref{5.26}---note that the sign of~$q_-$ changes when~$q_2$ and~$q_3$ are interchanged. Now the first two terms in \eqref{5.26} is exactly the determinant of~$\Pi_1(\bk)$. Hence \eqref{Pieq} follows.

If we plug in any of the four linear factors in~$E_1,E_2,E_3$ on the right-hand side of \eqref{Pieq} into the logarithm in \eqref{4.56}, integrate and apply the symmetries of the measure~$\textd\bk$, the result will be $\frac12 F(\theta^\star)$. Suppose now that~$\theta^\star\ne0^\circ,180^\circ$ and note that this implies that~$q_2\ne q_3$. Then~$q_2E_2+q_3E_3\ne q_3E_2+q_2E_3$ on a set of positive measure~$\textd\bk$. Hence, the two terms on the right-hand side of \eqref{Pieq} are not equal almost surely which by the strict concavity of the logarithm and Jensen's inequality implies that $\widetilde F_1(\theta^\star)>F(\theta^\star)$. But both functions are continuous in~$\theta^\star$, and so $\widetilde F_1(\theta^\star)-F(\theta^\star)$ is uniformly positive on any compact subset of the unit circle not containing~$0^\circ$ and~$180^\circ$. The existence of a desired~$c_3$ follows.
\end{proofsect}

\section{Probabilities of bad events}
\label{sec6}\noindent
Our goal here is to prove the estimate in Theorem~\ref{thm3.1} concerning the probability of a simultaneous occurrence of several bad events. While some of the details may still appear to be rather intricate, the principal input into the forthcoming argument has already been established in Sects.~\ref{sec4}--\ref{sec5}.

\subsection{Reflection positivity and chessboard estimates}
\label{sec3.1}\noindent
In this section we will glean from the classic theory of \emph{reflection positivity} those items that are needed at hand. Recall our notation~$\BbbP_{L,\beta}$ for the Gibbs probability measure on~$\T_L$ defined by the Hamiltonian~\eqref{Hdiff} at inverse temperature~$\beta$. Reflection positivity is a property of the measure~$\BbbP_{L,\beta}$ with respect to reflections of the torus which are defined as follows: Suppose that~$L$ is even and let us split the torus symmetrically into the ``left'' and ``right'' parts,~$\T_L^-$ and~$\T_L^+$, such that the two reflection-symmetric halves either share two planes of sites (reflections ``through sites'') or not (reflections ``through bonds''). Let~$P$ be the formal notation for the ``plane of reflection'' and let~$\scrF_P^+$, resp.,~$\scrF_P^-$ denote the~$\sigma$-algebra of events that depend only on the portion of the configuration in~$\T_L^+$, resp.,~$\T_L^-$. Introduce the reflection operator~$\vartheta_P$ on configurations in~$\T_L$, which induces a corresponding map $\vartheta_P\colon\scrF_P^+\to\scrF_P^-$. Then we have:

\begin{lemma}[Reflection positivity]
Consider the plane~$P$, the~$\sigma$-algebra~$\scrF_P^+$ and the measure~$\BbbP_{L,\beta}$ as specified above. Let~$\E_{L,\beta}$ denote the expectation with respect to~$\BbbP_{L,\beta}$. Then the following holds for all bounded $\scrF_P^+$-measurable random variables~$X$ and~$Y$:
\begin{equation}
\E_{L,\beta}\bigl(X\vartheta_P(Y)\bigr)=\E_{L,\beta}\bigl(Y\vartheta_P(X)\bigr)
\end{equation}
and
\begin{equation}
\label{pos}
\E_{L,\beta}\bigl(X\vartheta_P(X)\bigr)\ge0.
\end{equation}
Here~$\vartheta_P(Y)$ denotes the random variable~$Y\circ\vartheta_P$, and similarly for~$\vartheta_P(X)$.
\end{lemma}

\begin{proofsect}{Proof}
This is the standard reflection positivity proved in~\cite{FL,FILS1,FILS2}, which for reflections ``through sites'' follows simply by the fact that the interaction is exclusively via nearest neighbors, while for reflections ``through bonds'' it follows from this and the fact that the coupling is both quadratic and attractive.
\end{proofsect}

\begin{remark}
We remark that in the present work we use only the more robust version of reflection positivity---poor man's RP---which only requires nearest-neighbor interactions. (An exception to this ``rule'' is perhaps the argument leading to \eqref{4.26a}; but there we also offer an alternative approach via \cite{BCG}.)
\end{remark}

Our use of reflection positivity will come through the so called \emph{chessboard estimates}. To motivate the forthcoming definitions, let us briefly recall the principal idea. Using the expression on the left-hand side of \eqref{pos}, one can define an inner product on the~$\scrF_P^+$-measurable functions, which then satisfies the Cauchy-Schwarz inequality
\begin{equation}
\BbbP_{L,\beta}\bigl(\AA\cap\vartheta_P(\AA')\bigr)^2\le
\BbbP_{L,\beta}\bigl(\AA\cap\vartheta_P(\AA)\bigr)
\BbbP_{L,\beta}\bigl(\AA'\cap\vartheta_P(\AA')\bigr),
\end{equation}
for any~$\AA,\AA'\in\scrF_P^+$.
The interpretation of this inequality is that two given events, one on the ``left'' and the other on the ``right'' of~$\T_L$, can be separated within the expectation at the cost of reflecting both of them through~$P$. Iterating this bound further one can eventually disseminate each event all over the torus. The resulting quantity is often amenable to further analysis.

To state the chessboard estimates formally, let us consider a rectangular box~$V\subset\R^3$ of dimensions~$a_1\times a_2\times a_3$, where the~$a_i$'s are positive integers.  For simplicity, here and throughout this work, we assume that all of the~$a_i$ are related to~$L$ by powers of two, i.e., $a_i=2^{-m_i}L$ for some integers~$m_i$.
Consider the tiling of the (continuous) torus with dimensions~$L\times L\times L$ by translates of~$V$.  We will parametrize these translates by vectors $\bt\in\widetilde\T$ where~$\widetilde\T$ is the (discrete) torus with dimensions~$L/a_1\times L/a_2\times L/a_3$.

Let~$\AA$ be an event which depends only on configurations in~$V\cap\T_L$.  First we note that the event~$\AA$ can be reflected (multiply) through the various midplanes of~$V$, leading to seven new ostensibly different versions of the event~$\AA$.  [Labeling the resulting events by~$\sigma=(\sigma_1,\sigma_2,\sigma_3)\in\{0,1\}^3$, where~$\sigma_\alpha=1$ denotes whether the reflection in the~$\alpha$-th direction is implemented, we thus have altogether \emph{eight} events: one~$\AA_{000}=\AA$, three order-1 reflections $\AA_{100}$,~$\AA_{010}$ and~$\AA_{001}$ through the midplanes of~$V$ orthogonal to $x$,~$y$ and~$z$ lattice directions, respectively, three order-2 reflections~$\AA_{110},\AA_{101},\AA_{011}$ and one order-3 reflection~$\AA_{111}$.] Now if $\bt\in\widetilde\T$, let us define~$\vartheta_{\bt}(\AA)$, the appropriate notion of ``translation by $\bt$,'' as follows:
For~$\bt$'s with all even coordinates, this is simply the usual translation by $\bt$.  For~$\bt$'s with some odd coordinates, we select from the other seven versions of $\AA$ the one with reflections corresponding to all the odd coordinates of~$\bt$; the event $\vartheta_{\bt}(\AA)$ is then the translation by~$\bt$ of \emph{that} version of~$\AA$.

Let $Z_{L,\beta}(\AA)$ denote the constrained partition function defined by
\begin{equation}
Z_{L,\beta}(\AA)=Z_{L,\beta}\,\Bigl\langle\,\,\prod_{\bt\in\widetilde\T}
\1_{\vartheta_{\bt}(\AA)}\Bigr\rangle_{L,\beta},
\end{equation}
where~$Z_{L,\beta}$ is the usual partition function on~$\T_L$ and~$\1_{\vartheta_{\bt}(\AA)}$ denotes the indicator function of event~$\vartheta_{\bt}(\AA)$.
We are now ready for:

\begin{theorem}[Chessboard estimate]
\label{thm6.1}
Let the events $\AA_1,\dots,\AA_m$ and the partition functions $Z_{L,\beta}$ and $Z_{L,\beta}(\AA_k)$ be as above and let\/\/ $\bt_1,\dots,\bt_m$ be distinct vectors of the type described. Then
\begin{equation}
\label{CE}
\BbbP_{L,\beta}\Bigl(\,\bigcap_{k=1}^m\vartheta_{\bt_k}(\AA_k)\Bigr)
\le\prod_{k=1}^m\Bigl(\frac{Z_{L,\beta}(\AA_k)}{Z_{L,\beta}}\Bigr)^{1/|\widetilde\T|},
\end{equation}
where~$|\widetilde\T|$ is the volume of the factor torus~$\widetilde\T$.
\end{theorem}

\begin{proofsect}{Proof}
This is the standard \emph{chessboard estimate} proved in~\cite{FL,FILS1,FILS2}, see also \cite{Senya}. These estimates follow, in general, whenever the interaction is \emph{reflection positive}---here using both reflections through sites and bonds depending on whether the corresponding coordinate of the side of~$V$ is integer of half-integer.
\end{proofsect}

\begin{remark}  
We observe that due to the self-imposed evenness constraint on the dimensions of the tori, the objects $Z_{L,\beta}(\AA)$ and $Z_{L,\beta}(\vartheta_{\bt}(\AA))$ are identical for all $\bt$.  This will reduce the need for various provisos in the future derivations.
\end{remark}

In the forthcoming derivations the estimate \eqref{CE} will be used to bound the probability of a single bad event but, more importantly, to decouple various bad events. However, it will not be always advantageous to estimate~$Z_{L,\beta}(\AA_k)$ directly---often we will have to further decompose~$\AA_k$ into smaller events. Then we will use the well-known subadditivity property:

\begin{lemma}[Subadditivity]
\label{lemma7.3}
Consider the events~$\AA$ and~$(\AA_k)_{k\in\scrK}$ that depend only on configurations in a box~$V\cap\T_L$ where~$V$ is as specified above. If~$\AA\subset\bigcup_{k\in\scrK}\AA_k$, then
\begin{equation}
\label{subbound}
Z_{L,\beta}(\AA)\le \biggl(\,\sum_{k\in\scrK}Z_{L,\beta}(\AA_k)^{1/|\widetilde\T|}\biggr)^{|\widetilde\T|}.
\end{equation}
\end{lemma}

\begin{proofsect}{Proof}
The claim is (presumably) standard; we provide a short proof for reader's convenience. Clearly,~$Z_{L,\beta}(\AA)$ is equal to the~$Z_{L,\beta}$-multiple of the expectation of~$\prod_{\bt\in\widetilde\T}\1_{\vartheta_{\bt}(\AA)}$. Now, using the bound~$\1_{\vartheta_{\bt}(\AA)}\le\sum_{k\in\scrK}\1_{\vartheta_{\bt}(\AA_k)}$ we get
\begin{equation}
\label{6.7}
Z_{L,\beta}(\AA)\le Z_{L,\beta}\sum_{(k_{\bt})}\,\Bigl\langle\,\,\prod_{\bt\in\widetilde\T}
\1_{\vartheta_{\bt}(\AA_{k_{\bt}})}\Bigr\rangle_{L,\beta},
\end{equation}
where the collection $(k_{\bt})_{\bt\in\widetilde\T}\in\scrK^{\widetilde\T}$ provides the assignment of a $k_{\bt}\in\scrK$ for each of the translates of the $\AA$-events and where the sum is over all such assignments.
Applying Theorem~\ref{thm6.1},~$Z_{L,\beta}$ times the expectation on the right-hand side of \eqref{6.7} is bounded by the product of $Z_{L,\beta}(\AA_{k_{\bt}})^{1/|\widetilde\T|}$ over all~$\bt\in\widetilde\T$. But then each~$k_{\bt}$ can be independently summed over whereby the desired relation \eqref{subbound} follows.
\end{proofsect}

\subsection{Distinct types of badness}
\label{sec6.2a}\noindent
The estimate of the probability of bad events---defined right after Definition~\ref{def1} in Sect.~\ref{sec3.2}---will require partitioning this event into further categories.
\emph{A priori}, we will distinguish two types of badness according to which violation of the aforementioned conditions in Definition~\ref{def1} is highlighted. Specifically, we define the events
\begin{equation}
\label{BBE}
\BBE=\bigl\{\bS\colon |S^{(\alpha)}_{\br}-S^{(\alpha)}_{\br+\hate_\alpha}|\ge\Gamma
\text{ for some }\br,\br+\hate_\alpha\in\Lambda_B\bigr\}
\end{equation}
and
\begin{equation}
\label{BBSW}
\BBSW=\BB\setminus\BBE.
\end{equation}
Here~$\Lambda_B$ is the cube of~$(B+1)^3$ sites with the ``lowest left-most'' site at the origin (this is where the prototype bad event~$\BB$ was defined).
The idea behind this splitting is that for the configurations in~$\BBE$ there is an energetic ``disaster'' while for those in~$\BBSW$ the spin-wave approximation is still good but we are not particularly close to any free-energy minimum.

Unfortunately, the event~$\BBSW$ is still too complex to be estimated directly because after dissemination all over the torus, the resulting partition function~$Z_{L,\beta}(\BBSW)$ does not end up being the type featured in Sects.~\ref{sec4}--\ref{sec5}.  This is directly related to the fact---eluded to earlier---that there are a myriad of ground states in these models.  Thus we will have to work a bit in order to parcel~$\BBSW$ into events which after eventual dissemination over the torus lead to partition functions of a type discussed in the previous sections. 

\smallskip
In order to motivate the forthcoming definitions, let us categorize, somewhat more precisely than in Sect.~\ref{sec1.3}, the ground states of the model \eqref{Hdiff}. To avoid intricacies due to boundary conditions, we will restrict ourselves to toroidal geometry. First, all constant spin configurations minimize the energy. Second, more ground states can be generated from a homogeneous configuration by picking a lattice direction,~$\alpha$, and a sequence of planes orthogonal to~$\hate_\alpha$, and reflecting all spins in these planes through~$\alpha$-th of the vectors~$\hata,\hatb,\hatc$. These statements are more or less fully justified by Proposition~\ref{prop7.3} below. Of course, when we split~$\BBSW$ into the homogeneous and inhomogeneous parts, we will not try to keep track of all planes of reflection that can occur---one will be sufficient (this is the basis of the event~$\BB_{\alpha,j}^{(i)}$ below).

The decomposition of the event~$\BBSW$ will involve all of our basic scales: For a given~$\kappa>0$, $\Gamma>0$ and~$B>0$ we let
\begin{equation}
\label{DG}
\aD=\frac{12B\Gamma}\kappa.
\end{equation}
(As we will see in Sect.~\ref{sec6.2}, this will be the~$\aD$ for which we will use the results of spin-wave analysis from Sect.~\ref{sec4}.)
Let us parametrize the spins using the angle variables~$\theta_{\br}$. Fix an integer $s>1$ and let $\theta_1^\star,\dots,\theta_s^\star$ be~$s$ points uniformly spaced on the unit circle. The first part we want to identify from~$\BBSW$ are the nearly homogenous configurations: For each~$i\in\{1,\dots,s\}$, let~$\BB_0^{(i)}$ denote the event that the block~$\Lambda_B$ is bad and that
$|\theta_\br-\theta_i^\star|<\aD$ holds for all~$\br\in\Lambda_B$. 

The complementary part of~$\BBSW$ will feature a particular kind of inhomogeneity: Fix an $\alpha\in\{1,2,3\}$ and let~$j\in\{1,2,\dots,B\}$ and let~$\BbbH_j$ denote the plane in~$\T_L$ where all~$\br$ have the~$\alpha$-th component equal to~$j$. Fix~$i\in\{1,\dots,s\}$. If the angle between~$\theta_i$ and the~$\alpha$-th of the vectors~$\hata,\hatb,\hatc$ is within~$(-\kappa,\kappa)$ or~$(\pi-\kappa,\pi+\kappa)$, then we set~$\BB_{\alpha,j}^{(i)}=\emptyset$. For the other~$i$ we let $\BB_{\alpha,j}^{(i)}$ denote the set of all configurations~$\bS\in\BBSW$ such that $|\theta_{\br}-\theta_i^\star|<\aD$ holds for all~$\br\in\Lambda_B\cap\BbbH_{j-1}$ and $|\theta_{\br}-\tilde\theta_i^\star|<\aD$ for all $\br\in\Lambda_B\cap\BbbH_j$. Here~$\tilde\theta_i^\star$ denotes the angle~$\theta_i^\star$ reflected through the~$\alpha$-th of the vectors~$\hata$, $\hatb$ and~$\hatc$.

\begin{remark}
Let us reiterate that, \emph{by definition}, we have~$\BB_0^{(i)}\subset\BB$ for all~$i$ and $\BB_{\alpha,j}^{(i)}=\emptyset$ for all~$i$ whose~$\theta_i$ is too ``near'' the $\alpha$-th of the vectors~$\hata,\hatb,\hatc$. These facts will be useful when we estimate the associated partition functions~$Z_{L,\beta}(\BB_0^{(i)})$ and $Z_{L,\beta}(\BB_{\alpha,j}^{(i)})$ in Sect.~\ref{sec6.2}.
\end{remark}

It remains to show that the union of these events contains all of~$\BBSW$:

\begin{theorem}
\label{thm7.4}
Consider the~120$^\circ$-model and let the events~$\BBSW$, $\BB_0^{(i)}$ and $\BB_{\alpha,j}^{(i)}$ be as defined above. Suppose that~$\Gamma$,~$\kappa$,~$B$ and~$s$ are such that~$B\sqrt\Gamma\ll\kappa\ll1$ and $s\aD>4\pi$. Then
\begin{equation}
\label{BSWdec}
\BBSW\subseteq\bigcup_{i=1}^s\Bigl(\,\BB_0^{(i)}\cup\!\!\bigcup_{\alpha=1,2,3}\,\bigcup_{j=1}^B\BB_{\alpha,j}^{(i)}\Bigr).
\end{equation}
\end{theorem}

\begin{remark}
In the above and in what is to follow (and to a certain extent retroactively) we employ the symbol~``$\ll$'' in our hypotheses according to the standard fashion: ``if~$a\ll b$\dots'' means ``if the ratio~$a/b$ is bounded by a sufficiently small numerical constant which is uniform in any of the other parameters mentioned\dots.'' 
\end{remark}

\begin{remark}
The inclusion \eqref{BSWdec} justifies our previous claim that the only spin-wave calculations we need to do are against either homogeneous or stratified background. Indeed, by Lemma~\ref{lemma7.3}, to estimate the probability of~$\BBSW$ we will only need to estimate the constrained partition functions~$Z_{L,\beta}(\BB_0^{(i)})$ and $Z_{L,\beta}(\BB_{\alpha,j}^{(i)})$. The former leads directly to homogeneous spin-wave calculations from Sect.~\ref{sec4.1}; the latter will require further disemination of the pair of $(j-1,j)$-th planes in the~$\alpha$-direction which results in exactly the stratified background configuration treated in Sect.~\ref{sec4.3}. See Lemmas~\ref{lemma5.8} and~\ref{lemma5.9} for details.
\end{remark}

The proof of Theorem~\ref{thm7.4} commences by considering an elementary cube in~$\T_L$, say~$\K=\{0,1\}^3$, and classifying all spin configurations on~$\K$ that are ``nearly'' a ground state but which are not near any of the six ``priviledged'' directions $\hatw_\tau$; see \eqref{2.1} for the corresponding definition. The precise statement is as follows:

\begin{proposition}
\label{prop7.3}
Let~$\Gamma$ and~$\kappa$ be such that~$\sqrt\Gamma\ll\kappa\ll1$.
Let\,~$\btheta=(\theta_{\br})$ be a configuration of angle variables on~$\K$ such that the corresponding spins~$\bS_{\br}$ satisfy the energy constraints \eqref{3.1} for all pairs of nearest neighbors on~$\K$ but such that not all of the spins are within angle~$\kappa$ of \emph{one} particular~$\hatw_\tau$. Let~$\br\in\K$. Then (exactly) one of the following is true:
\settowidth{\leftmargini}{(111)}
\begin{enumerate}
\item[(1)]
$|\theta_{\br'}-\theta_{\br}|<4\Gamma/\kappa$ for all~$\br'\in\K$.
\item[(2)]
There exists an~$\alpha\in\{1,2,3\}$ such that $|\theta_{\br'}-\theta_{\br}|<4\Gamma/\kappa$ holds for all~$\br'\in\K$ with $\br-\br'\perp\hate_\alpha$, while for the remaining~$\br\in\K$ we have $|\theta_{\br'}-\tilde\theta_{\br}|<4\Gamma/\kappa$, where~$\tilde\theta_{\br}$ is obtained from~$\theta_{\br}$ by reflection through the~$\alpha$-th of the vectors~$\hata,\hatb,\hatc$.
\end{enumerate}
\end{proposition}

\begin{remark}
\label{rem9}
Setting~$\Gamma=0$ (and~$\kappa=0$) in this statement justifies Fig.~\ref{fig1}, which shows four examples of ground state configurations on an elementary cube. The reason why we explicitly exclude the ``almost'' constant configurations which point near one of~$\hatw_\tau$ is that, for these situations, the energy constraint would permit fluctuations that are of order~$\sqrt\Gamma$.
\end{remark}

The proof of Proposition~\ref{prop7.3} will involve a couple of lemmas. First, let us characterize the consequence of the energy constraint \eqref{3.1} for a single bond:

\begin{lemma}
\label{lemma7.4}
Let~$\alpha\in\{1,2,3\}$ and consider a nearest-neighbor bond~$(\br,\br')$ parallel to~$\hate_\alpha$. Let~$\theta_{\br}$ and~$\theta_{\br'}$ be two angle variables such that the corresponding spins satisfy
$|S_{\br}^{(\alpha)}-S_{\br'}^{(\alpha)}|<\Gamma$. Then either $|\theta_{\br}-\theta_{\br'}|<\pi\sqrt{\Gamma/2}$ or $|\theta_{\br}-\tilde\theta_{\br'}|<\pi\sqrt{\Gamma/2}$, where~$\tilde\theta_{\br'}$ is obtained from~$\theta_{\br'}$ by reflection through the~$\alpha$-th of the vectors~$\hata,\hatb,\hatc$.
\end{lemma}

\begin{proofsect}{Proof}
Without loss of generality, we will assume that~$\alpha=1$. Now, if~$\theta,\theta'\in[0,\pi]$ are two angles with~$|\theta-\theta'|=\epsilon$, then the trig identity~$|\cos\theta-\cos\theta'|=2|\sin(\tfrac{\theta-\theta'}2)|\sin(\tfrac{\theta+\theta'}2)$ and some optimization show that
\begin{equation}
|\cos\theta-\cos\theta'|\ge2\sin^2(\tfrac\epsilon2)\ge2\epsilon^2/\pi^2,
\end{equation}
where we used that~$\epsilon/2\in[0,\pi/2]$.
But the left hand side is exactly $|S_{\br}^{(1)}-S_{\br'}^{(1)}|$ which by assumption is less than~$\Gamma$. A simple algebra now shows that then~$\epsilon=|\theta-\theta'|\le\pi\sqrt{\Gamma/2}$. This proves the claim in the case when both~$\theta$ and~$\theta'$ have the same sign; the opposite case is handled by reflection through the~$x$ axis.
\end{proofsect}

Next we will extend this to a similar control of lattice plaquettes:

\begin{lemma}
\label{lemma7.5}
Let~$\BbbL$ be a lattice plaquette and let~$C=6\pi/\sqrt2$.  Let~$\Gamma\ll1$ and let~$\theta_{\br'}, \br'\in\BbbL$ denote angle variables such that the energy constraint~\eqref{3.1} holds for all four bonds.  Then, for any particular $\br\in\BbbL$, 
either all~$\theta_{\br'}$,~$\br'\in\BbbL$, are within~$C\sqrt\Gamma$ from~$\theta_{\br}$ or one neighbor of~$\br$ satisfies this condition while, on the other side of the plaquette, the other two spins are within~$C\sqrt{\Gamma}$ from the corresponding reflection of~$\theta_{\br}$.
\end{lemma}

\begin{proofsect}{Proof}
The proof is based on Lemma~\ref{lemma7.4}. To make the reference to this lemma easier, let us say that a \emph{reflection} occurs for the pair~$(\br,\br')$ if the latter possibility in Lemma~\ref{lemma7.4} applies. 
Let~$\BbbL$ be a lattice plaquette. Since permutations of coordinate directions can be matched with permuting the roles of~$\hata,\hatb,\hatc$, we can as well assume that~$\BbbL$ is an~$xy$-plaquette, i.e., $\BbbL=\{\br,\br+\hate_x,\br+\hate_y,\br+\hate_x+\hate_y\}$.
The analysis proceeds by checking various cases of increasing level of complexity. To simplify the formulas, let us abbreviate the error constant from Lemma~\ref{lemma7.4} by $\zeta=\pi\sqrt{\Gamma/2}$.

\smallskip\noindent
\textsl{CASE 1}:
\emph{No reflection occurs for both of the bonds emanating from~$\br$}. Lemma~\ref{lemma7.4} then implies that both~$\theta_{\br+\hate_x}$ and~$\theta_{\br+\hate_y}$ are within~$\zeta$ from~$\theta_{\br}$.   Now if a reflection does not occur on either of the two remaining bonds, then the spin at $\br+\hate_x+\hate_y$ is within $2\zeta$ of $\theta_{\br}$ and we are done.  The remaining possibility would be a reflection on both of these bonds.  But then~$\theta_{\br+\hate_x+\hate_y}$ is within~$2\zeta$ of both~$-\theta_{\br}$ and~$(-\tfrac{2\pi}3-\theta_{\br})$ which is impossible once~$4\zeta<\tfrac{2\pi}3$.

\smallskip\noindent
\textsl{CASE 2}: 
\emph{Reflection occurs for exactly one bond emanating from~$\br$}, say the horizontal bond from~$\br$. The only case we need to consider is when reflection occurs for the ``other'' vertical bond and does not for the ``other'' horizontal bond.  But then~$\theta_{\br+\hate_x+\hate_y}$ is within~$2\zeta$ of both~$\theta_{\br}$ and~$(-\tfrac{2\pi}3+\theta_{\br})$ which is again impossible once~$4\zeta<\tfrac{2\pi}3$.

\smallskip\noindent
\textsl{CASE 3}: 
\emph{Reflection occurs for both bonds emanating from~$\br$}. Clearly, following the path through $\br+\hate_x$ tell us that~$\theta_{\br+\hate_x+\hate_y}$ is within~$2\zeta$ of either $-\theta_{\br}$ or~$(-\tfrac{2\pi}3+\theta_{\br})$ while the passage through~$\br+\hate_y$ tells us that~$\theta_{\br+\hate_x+\hate_y}$ is within~$2\zeta$ of~$\pm(\tfrac{2\pi}3+\theta_{\br})$. Checking the cases shows that, if~$4\zeta<\tfrac{2\pi}3$, this is only possible when reflection occurs for \emph{all} bonds around the plaquette and when~$\theta_{\br}$ is within~$2\zeta$ of one of the angles~$0$,~$\tfrac\pi3$,~$\pi$ or~$\tfrac{4\pi}3$. Let us check the case when~$\theta_{\br}\approx0$. Then~$\theta_{\br+\hate_x}$ is within~$5\zeta$ of~$\theta_{\br}$ and both~$\theta_{\br+\hate_y}$ and~$\theta_{\br+\hate_x+\hate_y}$ are within~$6\zeta$ from~$(-\tfrac{2\pi}3-\theta_{\br})$. A similar argument handles the remaining cases.

\smallskip
Inspecting the above derivations, we see that the worst-case fluctuation from one of the two situations described in the statement of the lemma is by~$6\zeta=C\sqrt\Gamma$. This finishes the proof.
\end{proofsect}

\smallskip
Now we are ready to characterize the ``near'' ground states on elementary cubes:

\begin{proofsect}{Proof of Proposition~\ref{prop7.3}}
Lemma~\ref{lemma7.5} immediately implies that any configuration~$\bS$ satisfying the energy constraints~\eqref{3.1} on~$\K$ is one of the types featured in the statement of the proposition (resp.,~ Fig.~\ref{fig1}) to within errors~$C'\sqrt\Gamma$ for some numerical constant~$C'$. Indeed, either all angle variables are within~$C'\sqrt\Gamma$ of some particular angle or not. If not, then there must be a pair of nearest neighbors~$(\br,\br')$, say parallel to~$\hate_1$, where a reflection occurred. Then~$\theta_{\br'}$ is within~$C\sqrt\Gamma$ of~$-\theta_{\br}$. Moreover, choosing~$C'\gg C$ allows us to assume that both~$|\theta_{\br}|$ and~$|\pi-\theta_{\br}|$ exceed~$2C\sqrt\Gamma$ and thus both plaquettes in~$\K$ containing the bond~$(\br,\br')$ will have to be of a ``mixed'' type. But, again by Lemma~\ref{lemma7.5}, the two perpendicular $yz$-plaquettes cannot be of a ``mixed'' type. This implies the characterization in the statement of the proposition with the errors bounded by~$C'\sqrt\Gamma$.

It remains to show that the errors are in fact only proportional to~$\Gamma$ (cf Remark~\ref{rem9}). Here we will use the following refinement of Lemma~\ref{lemma7.4}: If~$\theta,\theta'\in[0,\pi)$ satisfy the energy constraint~$|\cos\theta-\cos\theta'|<\Gamma$ but are not within angle~$\kappa\ll1$ of the~$x$-ground states, then
\begin{equation}
\label{refin}
|\theta-\theta'|<\Gamma/\kappa.
\end{equation} 
Indeed, the Mean Value Theorem gives us that~$|\cos\theta-\cos\theta'|=2\sin(\theta'')(\theta-\theta')$ where~$\theta''$ lies between~$\theta$ and~$\theta'$. Hence~$\sin(\theta'')\ge\sin\kappa$ which for~$\kappa\ll1$ exceeds~$\kappa/2$. Using that~$|\cos\theta-\cos\theta'|<\Gamma$, the bound \eqref{refin} directly follows.

The improved error bound is now a simple consequence of \eqref{refin}. 
Let us first consider the ``nearly'' homogeneous situations. Since all angle variables are to be away from the ground state, \eqref{refin} implies that for each bond the $\theta_{\br}$'s will differ only by at most~$\Gamma/\kappa$. Hence, all $\theta_{\br}$'s on the cube must be within~$3\Gamma/\kappa$ of one of them which proves the claim in this case. The ``mixed'' configurations whose both types point away from any of the ground states are handled analogously, so we only have to consider the case when each type is within angle~$2\kappa$ of a different ground state. A generic situation of this kind is when the ``bottom''~$xy$-plaquette of~$\K$ is occupied by a configuration~$\theta\approx0$ while the ``top''~$xy$-plaquette is occupied by a configuration~$\theta\approx\tfrac{2\pi}3$. Then the observation \eqref{refin} constrains the size of the fluctuations to less than~$\Gamma/\kappa$ along the following bonds: The~$\hate_y$-bonds in the ``bottom''~$xy$-plaquette, all of the vertical bonds and the~$\hate_x$-bonds in the ``top''~$xy$-plaquette. It is easy to check that, from~$\br$, one can get to all sites of~$\K$ in at most \emph{four} steps, so all~$\theta_{\br'}$ are within less than~$4\Gamma/\kappa$ of~$\theta_{\br}$ or the corresponding reflection.
\end{proofsect}

\smallskip
Finally, we are ready to prove Theorem~\ref{thm7.4}:

\begin{proofsect}{Proof of Theorem~\ref{thm7.4}}
Consider a spin configuration~$\bS$ on~$\Lambda_B$ such that~$\BBSW$ occurs and let~$\theta_{\br}$ be the corresponding angle variables. Suppose first that one of the~$\theta_\br$'s makes an angle at least~$2\kappa$ with all of the~$\hatw_\tau$, $\tau=1,\dots,6$. Applying Lemma~\ref{lemma7.4} along with the fact that~$B\sqrt\Gamma\ll\kappa$, we find out that \emph{all} $\theta_{\br}$'s will be make an angle at least~$\kappa$ with any of the~$\hatw_\tau$. Proposition~\ref{prop7.3} then guarantees that any elementary cube has a layered structure with the~$\theta_{\br}$'s more or less constant in both layers. Since the maximal fluctuation in each elementary cube is at most~$4\Gamma/\kappa$, it is not more than~$3B$-times that---i.e.,~$\aD$ in \eqref{DG}---for any pair of spins~in~$\Lambda_B$. 

Now the bound~$s\aD>4\pi$ ensures that the consecutive~$\theta_i^\star$ (which we used to define the events $\BB_0^{(i)}$ and $\BB_{\alpha,j}^{(i)}$) are within less than~$\aD/2$ from each other. Thus, if all spins point in about the same direction they must all be within~$\aD$ of some~$\theta_i^\star$---which implies that~$\bS\in\BB_0^{(i)}$---or there are two consecutive layers, say~$j-1$ and~$j$, in the~$\alpha$-th lattice direction where a reflection from~$\theta_i^\star$ to~$\tilde\theta_i^\star$ occurs. In the latter case we have~$\bS\in\BB_{\alpha,j}^{(i)}$. This proves \eqref{BSWdec} for those~$\bS\in\BBSW$ for which at least one of the spins is farther than~$2\kappa$ (in the angular distance) from any of the six preferred directions~$\hatw_\tau$.

It remains to deal with the situations in which \emph{all} spins are within~$2\kappa$ of some $\hatw_\tau$ (possibly different~$\tau$ for different spins). Clearly, the latter cannot be the same for all spins because of the inclusion $\BBSW\subset\BB$, and so there must be a pair of spins where the type of ground state is different at the endpoints. But then we can still use Proposition~\ref{prop7.3} for the elementary cubes containing this bond, and then the cubes next to these and so on. In this way we conclude that the endpoints of this bond belong to two parallel planes of sites where the spins do not fluctuate by more than~$2B$ times~$4\Gamma/\kappa$ about a single direction in one plane and its reflection in the other. Hence~$\bS$ belongs to one of the $\BB_{\alpha,j}^{(i)}$'s.
\end{proofsect}

\subsection{Proof of Theorem~\ref{thm3.1}}
\label{sec6.2}\noindent
We begin with an estimate of the partition function for event~$\BBE$.

\begin{lemma}
\label{lemma5.7}
Let $\kappa>$ be fixed. There exist constants~$c_4\in(0,\infty)$ and~$\delta>0$ such that if $\beta J$, $\aD=12B\Gamma/\kappa$ and~$\delta$ satisfy the bounds \eqref{4.3} then
\begin{equation}
\label{6.13a}
\limsup_{L\to\infty}\,
\Bigl(\frac{Z_{L,\beta}(\BBE)}{Z_{L,\beta}}\Bigr)^{(B/L)^3}\le B^3(c_4\beta J)^{B^3/2}
e^{-\frac12\beta J\Gamma^2}.
\end{equation}
\end{lemma}

\begin{proofsect}{Proof}
We will derive an upper bound~$Z_{L,\beta}(\BBE)$ and a lower bound on~$Z_{L,\beta}$. The former is essentially an immediate consequence of the definition of~$\BBE$. Indeed, on $\BBE$ at least one of the pairs of nearest neighbors in~$\Lambda_B$ contributes at least $\frac12(\beta J)\Gamma^2$ to the total energy. Thus, after dissemination of~$\BBE$ all over the torus, the spin configurations are constrained to satisfy
\begin{equation}
\label{6.15a}
\beta\mathscr{H}_L(\bS)\ge\frac12(\beta J)\Gamma^2 
\Bigl(\frac LB\Bigr)^3.
\end{equation}
It follows that
\begin{equation}
Z_{L,\beta}(\BBE)^{(B/L)^3}
\le 6B^3 (2\pi)^{B^3}e^{-\frac12\beta J\Gamma^2},
\end{equation}
where the factor $6B^3$ bounds the number of places where the ``excited'' bond can occur within~$\Lambda_B$ and~$(2\pi)^{B^3}$ is the total ``phase volume'' of all configurations in~$\Lambda_B$.

Next we need to derive a lower bound on~$Z_{L,\beta}$. Here we will write the partition function as an integral of~$e^{-\beta\mathscr{H}_L}$; a lower bound can then be obtained by inserting the indicator that all angle variables are within~$\Delta$ of~$0^\circ$. This yields
\begin{equation}
Z_{L,\beta}\ge\Bigl(\frac{2\pi}{\beta J}\Bigr)^{L^3/2}
e^{-L^3 F_{L,\beta}^{(\Delta)}(0^\circ)},
\end{equation}
where~$F_{L,\beta}^{(\Delta)}(0^\circ)$ is the quantity from \eqref{FDLB1}.
Choosing~$\epsilon>0$ and letting~$\delta$ be such that Theorem~\ref{thm4.1} holds, we thus get
\begin{equation}
\label{6.17a}
\liminf_{L\to\infty}\,(Z_{L,\beta})^{1/L^3}\ge \Bigl(\frac{2\pi}{\beta J}\Bigr)^{1/2}e^{-F(0^\circ)-\epsilon},
\end{equation}
where~$F$ denotes the spin-wave free energy \eqref{Fn1}. Combining \eqref{6.15a} and \eqref{6.17a} and letting~$c_4$ absorb all factors independent of~$B$ and~$\beta J$, the desired bound \eqref{6.13a} is proved.
\end{proofsect}

\begin{remark}
Since the event~$\BBE$ depends only on~$\Gamma$, the appearance of~$\kappa$ in the assumptions of Lemma~\ref{lemma5.7} may seem unnecessary. However, some conditions on~$\Gamma$, $B$ and~$\beta J$ are still needed to derive the lower bound in \eqref{6.17a} and the advantage of the present form is that now all lemmas in this section are proved under more or less the same assumptions.
\end{remark}

Next we will attend to the event~$\BBSW$. In light of Theorem~\ref{thm7.4} and Lemma~\ref{lemma7.3}, we can focus directly on the events~$\BB_0^{(i)}$ and~$\BB_{\alpha,j}^{(i)}$. We will begin with the former of the two:

\begin{lemma}
\label{lemma5.8}
Let~$\kappa>0$ be fixed. There exist numbers~$\rho_1(\kappa)>0$ and $\delta>0$ such that if~$\beta J$ and~$\aD=12B\Gamma/\kappa$ satisfy the bounds \eqref{4.3} with this~$\delta$ and if~$B\Gamma\ll\kappa\ll1$, then
\begin{equation}
\limsup_{L\to\infty}\,
\Bigl(\frac{Z_{L,\beta}(\BB_0^{(i)})}{Z_{L,\beta}}\Bigr)^{1/L^3}\le e^{-\rho_1(\kappa)},
\qquad i=1,\dots,s.
\end{equation}
\end{lemma}

\begin{proofsect}{Proof}
To summarize the situation, on~$\BB_0^{(i)}$, all angle variables~$\theta_{\br}$ in the block~$\Lambda_B$ are within $\aD$ of~$\theta_i^\star$. 
If we now consider the multiply reflected event associated with~$\BB_0^{(i)}$, the same will be true about all spins on~$\T_L$. Let~$\epsilon>0$ and let~$\delta>0$ be as in Theorem~\ref{thm4.1}. Then
\begin{equation}
\limsup_{L\to\infty}Z_{L,\beta}(\BB_0^{(i)})^{1/L^3}\le
\Bigl(\frac{2\pi}{\beta J}\Bigr)^{1/2}
e^{-F(\theta_i^\star)+\epsilon}.
\end{equation}
Using \eqref{6.17a} we thus conclude
\begin{equation}
\label{zzB0}
\limsup_{L\to\infty}\,
\Bigl(\frac{Z_{L,\beta}(\BB_0^{(i)})}{Z_{L,\beta}}\Bigr)^{1/L^3}\le e^{-F(\theta_i^\star)+F(0^\circ)+2\epsilon}.
\end{equation}
It remains to adjust~$\epsilon$ so that the exponent is negative. Here we first note that~$\BB_0^{(i)}$ is empty unless~$\theta_i^\star$ is at least~$\kappa$ away from any of the ground state (indeed, otherwise the configuration fails to be in~$\BB$, which by definition contains~$\BB_0^{(i)}$).
Applying Corollary~\ref{cor5.2}, $F(\theta^\star)$ exceeds~$F(0^\circ)$ by a uniformly positive amount, denoted by~$2\rho_1(\kappa)$, whenever~$\theta^\star$ is at least~$\kappa$ away from the minimizing angles. Now choose~$\epsilon\le\tfrac12\rho_1(\kappa)$ and let~$\delta$ be the corresponding quantity from Theorem~\ref{thm4.1}. Then the right-hand side of \eqref{zzB0} is indeed less than~$e^{-\rho_1(\kappa)}$, proving the desired claim.
\end{proofsect}

Similarly, we have to derive a corresponding bound for the events~$\BB_{\alpha,j}^{(i)}$:

\begin{lemma}
\label{lemma5.9}
Let~$\kappa>0$ be fixed. There exist numbers~$\rho_2(\kappa)>0$ and $\delta>0$ such that if~$\beta J$ and~$\aD=12B\Gamma/\kappa$ satisfy the bound \eqref{4.3s} with this~$\delta$ and if~$B\Gamma\ll\kappa\ll1$, then
\begin{equation}
\label{zzbd}
\limsup_{L\to\infty}\,
\Bigl(\frac{Z_{L,\beta}(\BB_{\alpha,j}^{(i)})}{Z_{L,\beta}}\Bigr)^{1/L^3}
\le e^{-\rho_2(\kappa)/B},
\end{equation}
holds for all~$\alpha\in\{1,2,3\}$, all~$j\in\{1,2,\dots,B\}$ and all~$i\in\{1,\dots,s\}$.
\end{lemma}

\begin{remark}
We assure the reader that the~$1/B$ in the exponent is no cause for alarm; in accord with \eqref{subbound}, the relevant object from Lemma~\ref{lemma5.9} is the right-hand side raised to power~$B^3$.
\end{remark}

\begin{proofsect}{Proof of Lemma~\ref{lemma5.9}}
Recall that, on~$\BB_{\alpha,j}^{(i)}$, all~$\theta_{\br}$ for~$\br$ in the plane~$\Lambda_B\cap\BbbH_{j-1}$ are within a constant times $B\Gamma/\kappa$ of~$\theta_i^\star$, while those in the neighboring plane~$\Lambda_B\cap\BbbH_j$ are within the same distance of the reflected angle~$\tilde\theta_i^\star$. After dissemination over the torus, which is what gives rise to the quantity~$Z_{L,\beta}(\BB_{\alpha,j}^{(i)})$, the same will be true about the spins in the \emph{entire} planes~$\BbbH_{j-1}$, resp.,~$\BbbH_j$, and also about their translates by integer multiples of~$B$ in the orthogonal direction. However, we cannot yet use the spin-wave calculation; instead, we have to use Theorem~\ref{thm6.1} again to disseminate the two-plane alternating pattern all over the torus. This yields
\begin{equation}
\label{zzB0a}
\frac{Z_{L,\beta}(\BB_{\alpha,j}^{(i)})}{Z_{L,\beta}}\le \biggl(\frac{Z_{L,\beta}(\widetilde\BB_{\alpha,j}^{(i)})}{Z_{L,\beta}}\biggr)^{2/B},
\end{equation}
where~$\widetilde\BB_{\alpha,j}^{(i)}$ is the event in~$\Lambda_B$ that the~$\theta_{\br}$ are within~$\Delta$ of~$\theta_i^\star$ in even translates of~$\BbbH_{j-1}$ and of~$\tilde\theta_i^\star$ in odd translates of~$\BbbH_{j-1}$. 

Now the partition function can be estimated using Theorem~\ref{thm4.10} and we thus get
\begin{equation}
\label{zzB1}
\limsup_{L\to\infty}\,
\biggl(\frac{Z_{L,\beta}(\widetilde\BB_{\alpha,j}^{(i)})}{Z_{L,\beta}}\biggr)^{1/L^3}
\le e^{-[\widetilde F_\alpha(\theta_i^\star)-F(0^\circ)-2\epsilon]},
\end{equation}
But Theorem~\ref{thm5.3} shows that~$\widetilde F_\alpha(\theta_i^\star)-F(0^\circ)\ge c_3>0$ for some~$c_3=c_3(\kappa)$ for all~$i$ for which~$\theta_i^\star$ is at least~$\kappa$-away from any of the minimizing angles associated with ``stratification'' direction~$\alpha$, while, by definition,~$\BB_{\alpha,j}^{(i)}=\emptyset$ for those~$i$ that fail this condition. 
Hence, if we choose~$\epsilon>0$ so small that~$\rho_2(\kappa)=2(c_3-2\epsilon)>0$ and let~$\delta$ be the corresponding constant from Theorem~\ref{thm4.10}, then \twoeqref{zzB0a}{zzB1} imply \eqref{zzbd} as desired.
\end{proofsect}

\begin{proofsect}{Proof of Theorem~\ref{thm3.1}}
Let $\kappa>0$ and let $\delta>0$ be the minimum of the corresponding numbers from Lemmas~\ref{lemma5.7}--\ref{lemma5.9}.
Fix an $\eta\in(0,1)$. We claim that for each sufficiently large~$\beta$, there exist numbers~$B$ and~$\Gamma$ such that the bounds \eqref{4.3} and \eqref{4.3s} for~$\aD=12B\Gamma/\kappa$ hold, the inequality~$B\sqrt\Gamma\ll\kappa$ can be achieved and the bound
\begin{equation}
\label{bd1}
B^3(c_4\beta J)^{B^3/2}
e^{-\frac12\beta J\Gamma^2}
+\frac{8\pi}{\Delta}e^{-B^3\rho_1(\kappa)}
+\frac{24\pi B}\Delta e^{-B^2\rho_2(\kappa)}
<\eta
\end{equation}
is true. Indeed, we can for instance take $B=\log\beta$ and $\Gamma=\beta^{-\frac5{12}}$ and note that, for these choices, the left-hand side will eventually decrease with~$\beta$.

Now choose~$s$ such that $s\aD>4\pi$ but~$s\aD<8\pi$. Then the definitions \twoeqref{BBE}{BBSW} of events~$\BBE$ and~$\BBSW$, the decomposition of~$\BBSW$ from Theorem~\ref{thm7.4}, the chessboard estimate and the (subadditivity) Lemma~\ref{lemma7.4} imply that~$\BbbP_{L,\beta}(\vartheta_{\bt_1}(\BB)\cap\dots\cap\vartheta_{\bt_m}(\BB))$ will be bounded by~$\eta_L^m$, where
\begin{multline}
\qquad
\eta_L=\Bigl(\frac{Z_{L,\beta}(\BBE)}{Z_{L,\beta}}\Bigr)^{(B/L)^3}
+\sum_{i=1}^s
\Bigl(\frac{Z_{L,\beta}(\BB_0^{(i)})}{Z_{L,\beta}}\Bigr)^{(B/L)^3}
\\+\sum_{i=1}^s\sum_{\alpha=1,2,3}\sum_{j=1}^B
\Bigl(\frac{Z_{L,\beta}(\BB_{\alpha,j}^{(i)})}{Z_{L,\beta}}\Bigr)^{(B/L)^3}.
\qquad
\end{multline}
By Lemmas~\ref{lemma5.7}--\ref{lemma5.9}, the fact that~$s<8\pi/\aD$ and~\eqref{bd1}, it follows that $\limsup_{L\to\infty}\eta_L<\eta$. Hence there exists a number~$L_0\in(0,\infty)$ such that~$\eta_L\le\eta$ for all~$L\ge L_0$. But for~$L\ge L_0$, the probability $\BbbP_{L,\beta}(\vartheta_{\bt_1}(\BB)\cap\dots\cap\vartheta_{\bt_m}(\BB))$ is bounded by~$\eta^m$ uniformly in~$m$ and the choice of the vectors~$\bt_1,\dots,\bt_m$. This proves the desired claim and thus also finishes the proof of our main result (Theorem~\ref{thm2.1}).
\end{proofsect}

\section{Spherical models}
\label{sec7}\noindent
Here we present the proof that the spherical version of the 120$^\circ$-model has no phase transition at any positive temperature. This demonstrates the failure of the naive spin-wave arguments and, particularly, the infrared bounds.

\smallskip
Spherical models, very popular in the 1950-60, were conceived of by Berlin-Kac~\cite{Berlin-Kac} as convenient approximations of the statistical mechanical systems which are more amenable to explicit computations.
(On the mathematics side, the topic received a new wave of interest in the~1980's through the rigorous versions of~$1/n$ expansion.) To construct a spherical version of a given spin system, we use the same Hamiltonian but ascribe different meaning to the spin variables. In particular, the local \emph{a priori} constraints on the spin variables are relaxed and are replaced by a global constraint. For instance, for the Ising model with Hamiltonian $\mathscr{H}=-\sum_{\br,\br'}\sigma_{\br}\sigma_{\br'}$ we have $\sigma_{\br}=\pm1$ and thus $\sigma_{\br}^2=1$ for all~$\br$. The spherical version has the same interaction Hamiltonian but now we only require that $(1/N)\sum_{\br}\sigma_{\br}^2=1$, where~$N$ denotes the total number of spins.

Often enough, these models are further simplified by stipulating that the constraint only needs to be satisfied in the \emph{mean} and may thus be enforced by Langrange multipliers. The latter type is often referred to as the \emph{mean spherical model}. This version usually turns out to be pretty much the same in most aspects, see~\cite{Gough-Pule} for some discussion. Here we will go the mean-spherical route partially because the resulting analysis is simpler, but also because the analogy to pure spin-wave theory is more pronounced in this case. We refer to~\cite[Section II.11]{Simon} for more references and further discussion.

Thus, we will take \eqref{1.11} as our basic Hamiltonian along with an additional term to enforce the required constraints. However now it is understood that the spin variables are no-longer constrained to the unit circle; the integration takes place over all of~$\R^2$. The constraining term reads $-\mu\sum_{\br,\alpha}(S_{\br}^{(\alpha)})^2$ but now (unfortunately) $S_{\br}^{(\alpha)}$ refers to the \emph{Cartesian} component of the spin. This means that we will have to rewrite the Hamiltonian in terms of the Cartesian components~of~$\bS$.

The key to the mean-spherical approximation is that for arbitrary~$\mu>0$ the partition function can be solved exactly by translations to spin-wave variables. Then $\mu$ is supposed to be adjusted so that the relevant constraint is enforced. As we shall see, if there is an infrared divergence, this adjustment is easy and everything is analytic in~$\beta$. In the opposite case, there may be a condensation at large~$\beta$ and if so, one may conclude---with a lot of apologies---that a phase transition has occurred. The primary conclusion of this section is that the latter possibility does \emph{not} materialize in the model at hand.

\smallskip
Now we are ready to describe the spherical version of the 120$^\circ$-model.
The Hamiltonian on torus~$\T_L$ is given by 
\begin{multline}
\label{Hsp2}
\quad
\beta\mathscr{H}_L=\frac{\beta J}2\sum_{\br\in\T_L}\biggl\{
\bigl(S_{\br}^{(x)}-S_{\br+\hate_x}^{(x)}\bigr)^2
+\Bigl[\bigl(\tfrac{\sqrt3}2S_{\br}^{(y)}-\tfrac12S_{\br}^{(x)}\bigl)-\bigl(\tfrac{\sqrt3}2S_{\br+\hate_y}^{(y)}-\tfrac12S_{\br+\hate_y}^{(x)}\bigl)\Bigl]^2
\\\qquad+\Bigl[\bigl(\tfrac{\sqrt3}2S_{\br}^{(y)}+\tfrac12S_{\br}^{(x)}\bigl)-\bigl(\tfrac{\sqrt3}2S_{\br+\hate_z}^{(y)}+\tfrac12S_{\br+\hate_z}^{(x)}\bigl)\Bigl]^2\biggr\},
\qquad
\end{multline}
where $S_{\br}^{(x)}$ and $S_{\br}^{(y)}$ are now unrestricted real variables \emph{a priori} distributed according to the Lebesgue measure on~$\R$. The constraint is represented by the quantity
\begin{equation}
\mathscr{N}_L=
\sum_{\br\in\T_L}\bigl((S_{\br}^{(x)})^2+(S_{\br}^{(y)})^2\bigl).
\end{equation}
The associated Gibbs measure is given in terms of the Radon-Nikodym derivative with respect to the Lebesgue measure on~$(\R^2)^{\T_L}$, which is simply a properly normalized~$e^{-\beta\mathscr{H}_L-\mu\mathscr{N}_L}$. We will denote the expectation with respect to the resulting thermal state by~$\langle-\rangle_{L,\beta,\mu}$.

\begin{theorem}
\label{thm7.2}
Consider the spherical 120$^\circ$-model with the Hamiltonian \eqref{Hsp2} and let $\langle-\rangle_{L,\beta,\mu}$ denote the corresponding thermal state for the chemical potential~$\mu$. Then there exists a positive, real-analytic function~$\mu_\star\colon[0,\infty)\to(0,\infty)$ such that for each~$\beta\in(0,\infty)$---and~$\mu$ set to~$\mu_\star(\beta)$---the following is true: The constraint is satisfied on average,
\begin{equation}
\label{7.2eq}
\lim_{L\to\infty}\frac1{L^3}\langle\mathscr{N}_L\rangle_{L,\beta,\mu_\star(\beta)}=1,
\end{equation}
there is no long range order,
\begin{equation}
\label{7.4}
\lim_{L\to\infty}\biggl\langle\Bigl|\frac1{L^3}\sum_{\br\in\T_L}\bS_{\br}\Bigr|^2\biggr\rangle_{L,\beta,\mu_\star(\beta)}=0,
\end{equation}
and the limiting measure exhibits a clustering property,
\begin{equation}
\label{7.5}
\lim_{|\br-\br'|\to\infty}
\lim_{L\to\infty}\langle S_{\br}^{(\alpha)}S_{\br'}^{(\alpha')}\rangle_{L,\beta,\mu_\star(\beta)}=0,
\end{equation}
for any~$\alpha,\alpha'\in\{x,y\}$.
Moreover, the limiting free energy is a (real) analytic function of~$\beta$.
\end{theorem}

\begin{proofsect}{Proof}
As usual, our first goal will be to calculate the limiting free energy as a function of~$\beta$ and~$\mu$. Let~$Z_L(\beta,\mu)$ denote the integral of $e^{-\beta\mathscr{H}_L-\mu\mathscr{N}_L}$ with respect to the Lebesgue measure on~$(\R^2)^{\T_L}$. In order to compute~$Z_L(\beta,\mu)$ we transform to the Fourier modes in which case the spin-wave Hamiltonian (including the constraint) is seen to be given by
\begin{multline}
\label{sphH}
\qquad
\beta\mathscr{H}_L+\mu\mathscr{N}_L=\frac{\beta J}2\sum_{\bk\in\T_L^\star}\biggl\{|\hatS_{\bk}^{(x)}|^2\bigl[E_1+\tfrac14(E_2+E_3)+\lambda\bigr]
\\+
|\hatS_{\bk}^{(y)}|^2\bigl[\tfrac34(E_2+E_3)+\lambda\bigr]
+\tfrac{\sqrt3}4\bigl(\hatS_{\bk}^{(x)}\hatS_{-\bk}^{(y)}+\hatS_{-\bk}^{(x)}\hatS_{\bk}^{(y)}\bigr)
[E_2-E_3]\biggr\}.
\qquad
\end{multline}
Here $\hatS^{(\alpha)}_{-\bk}$ is just the complex conjugate of~$\hatS^{(\alpha)}_{\bk}$, the symbol~$E_\alpha$ abbreviates the usual $E_\alpha(\bk)=|1-e^{\texti k_\alpha}|^2$ and~$\lambda$ is defined by $\beta J\lambda/2=\mu$. In terms of the two-component variable~$(\hatS_{\bk}^{(x)},\hatS_{\bk}^{(y)})$, the right-hand side (without the~$\beta J/2$ prefactor, of course) can be written as a quadratic form with matrix $\lambda\1+\Theta(\bk)$, where
\begin{equation}
\Theta(\bk)=
\left(\begin{matrix}
E_1+\tfrac14(E_2+E_3) &  \tfrac{\sqrt3}2(E_2-E_3)\\
\tfrac{\sqrt3}2(E_2-E_3) & \tfrac34(E_2+E_3)
\end{matrix}\right).
\end{equation}
In this notation the integrals are readily performed with the limiting free energy~$F(\beta,\lambda)$---which to within a sign is the limit~$\lim_{L\to\infty}L^{-3}\log Z_L(\beta,\beta J\lambda/2)$---given by
\begin{equation}
\label{7.4eq}
F(\beta,\lambda)=\log\frac{\beta J}{2\pi}+\frac12\int_{[-\pi,\pi]^3}\frac{\textd\bk}{(2\pi)^3}\log\det\bigr[\lambda\1+\Theta(\bk)\bigr].
\end{equation}
Here the integral converges as long as~$\lambda>0$.

Our next goal is to find the function~$\mu_\star$ for which \eqref{7.2eq} holds. Using standard relation between free energy and expectation, the constraint equation becomes
\begin{equation}
\label{derph}
\frac\partial{\partial\lambda}F(\beta,\lambda)=\frac{\beta J}2\int_{[-\pi,\pi]^3}\frac{\textd\bk}{(2\pi)^3}\CalS_{\text{SP}}(\bk)=\frac{\beta J}2,
\end{equation}
where
\begin{equation}
\label{7.10a}
\CalS_{\text{SP}}(\bk)=\lim_{L\to\infty}\langle|\bhatS_{\bk}|^2\rangle_{L,\beta,\mu}
=(\beta J)^{-1}\text{Tr}[\lambda\1+\Theta(\bk)]^{-1}
\end{equation}
is the so-called \emph{structure factor}.
As long as~$\lambda>0$, the derivative $\frac\partial{\partial\lambda}F(\beta,\lambda)$ is finite and independent of~$\beta$ and thus \eqref{derph} defines a function $\lambda\mapsto\beta_\star(\lambda)$. A moment's thought shows that this function is strictly decreasing and hence locally invertible. However, before we plug the inverse back into \eqref{7.4eq}, we need to establish the range of values that~$\beta_\star(\lambda)$ can take. In particular, we ask whether~$\beta_\star(\lambda)$ diverges as $\lambda\downarrow0$.

Examining the constraint equation in detail, the crucial issue boils down to convergence/diver\-gence of the momentum-space integral of the structure factor
\begin{equation}
\label{Eblah}
\CalS_{\text{SP}}(\bk)\propto\frac{E_1+E_2+E_3}{E_1E_2+E_1E_3+E_2E_3}.
\end{equation}
It turns out that the integral of~$\CalS_{\text{SP}}(\bk)$ \emph{diverges} although this is not apparent by naive power counting. Indeed, the primary source of the divergence is not the origin  but the coordinate axes. This is seen by an easy lower bound on~$\CalS_{\text{SP}}(\bk)$: Fix $k_3$ to a non-zero number and note that we can discard the~$E_1$ and~$E_2$ from the numerator. Second, the term $E_1E_2$ in the denominator is bounded above by a constant times $E_1+E_2$. Hence, the calculations boil down to the integral of $(E_1+E_2)^{-1}$ with respect to~$k_2$ and~$k_3$, which is manifestly divergent.

The above reasoning shows that $\lambda\mapsto\beta_\star(\lambda)$ takes all positive real values as~$\lambda$ sweeps through the positive real line and hence the inverse~$\beta\mapsto\lambda_\star(\beta)$ is defined for all~$\beta\in[0,\infty)$. Moreover, for~$\lambda>0$ the function $\lambda\mapsto\beta_\star(\lambda)$ is analytic in a small neighborhood of the real line and hence so is~$\beta\mapsto\lambda_\star(\beta)$. The desired function then arises by setting~$\mu_\star(\beta)=\beta J\lambda_\star(\beta)/2$, which satisfies \eqref{7.2eq} by construction. Furthermore, plugging~$\lambda_\star(\beta)$ for~$\lambda$ in~$F(\lambda,\beta)$ proves that the free energy is real analytic in~$\beta$. In order to prove also \twoeqref{7.4}{7.5}, we just need to note that \eqref{sphH} implies that the correlator~$\langle \hatS_{-\bk}^{(\alpha)}\hatS_{\bk}^{(\alpha')}\rangle_{L,\beta,\mu}$ is exactly the $(\alpha,\alpha')$-th matrix element of~$(\beta J)^{-1}[\lambda\1+\Theta(\bk)]^{-1}$. But~then
\begin{equation}
\label{7.4a}
\biggl\langle\Bigl|\frac1{L^3}\sum_{\br\in\T_L}\bS_{\br}\Bigr|^2\biggr\rangle_{L,\beta,\mu_\star(\beta)}=\frac2{\beta J\lambda_\star(\beta)L^3}\underset{L\to\infty}{\longrightarrow}0,
\end{equation}
while
\begin{equation}
\label{7.5a}
\lim_{L\to\infty}
\langle S_{\br}^{(\alpha)}S_{\br'}^{(\alpha')}\rangle_{L,\beta,\mu_\star(\beta)}
=\int_{[-\pi,\pi]^3}\frac{\textd\bk}{(2\pi)^3}\frac1{\beta J}
\Bigl(\frac1{\lambda_\star(\beta)\1+\Theta(\bk)}\Bigr)_{\alpha\alpha'}
e^{\texti\bk\cdot(\br-\br')},
\end{equation}
which by the Riemann-Lebesgue lemma and the fact that~$\lambda_\star(\beta)$ is strictly positive for any~$\beta\in[0,\infty)$ tends to zero as $|\br-\br'|\to\infty$.
\end{proofsect}

\begin{remark}
The last expression of the proof indicates that the correlations decay (at least) exponentially fast. However, as is seen from \eqref{Eblah}, the angular dependence of the resulting correlation length is fairly complicated. In particular, there may be directions in which the quadratic approximation of~$\Theta(\bk)$ vanishes in which case more than one pole in the ``complex~$|\bk|$'' plane (instead of the usual single pole) jointly contribute to the integral.
\end{remark}

We conclude with a remark concerning the relation of these findings to the actual systems of interest.
For the spherical model, the so called structure factor~$\CalS_{\text{SP}}(\bk)=\langle|\bhatS_{\bk}|^2\rangle_{\beta,\mu}$ can explicitly be computed, cf \eqref{7.10a}. As was established in \cite{FSS,FILS1,FILS2} for a general class of nearest-neighbor ferromagnetic systems (including the one discussed in the present work) the spherical rendition of the structure factor with~$\mu=0$ provides a bound on the structure factor~$\CalS_{\text{A}}(\bk)$ (namely, the two-point correlation function in~$\bk$-representation) of the \emph{actual} system, 
\begin{equation}
\CalS_{\text{A}}(\bk)\le\CalS_{\text{SP}}(\bk)\big|_{\mu=0}.
\end{equation}
This is the basis of the infrared-bound technology which uses the convergence of the integrated bound to establish long-range order at low temperatures.

Here, the low momentum behavior of the spherical structure factor together with the rigorous as well as non-rigorous results relating~$\CalS_{\text{SP}}$ to~$\CalS_{\text{A}}$ (including in particular~\cite{Harris}) strongly suggest a disordering due to long wave-length fluctuations. It is usually the case that these are reliable indicators for the behavior of the actual system. Evidently, as the results of this work show, the present cases are exceptional. 

\section*{Acknowledgments}
\noindent
The research of M.B. and L.C.~was supported by the NSF under the grant NSF~DMS-0306167. Parts of this paper were written when M.B.~was visiting Microsoft Research in Redmond whose hospitality is gratefully acknowledged. The authors wish to thank Jeroen van den Brink for discussions and clarifications and two anonymous referees for suggestions that led to improvements in the presentation.


\end{document}